\documentclass[preprint,3p,12pt,sort&compress]{elsarticle}

\usepackage{graphicx}
\usepackage{amsmath}
\usepackage{amsfonts}
\usepackage{enumitem} 							
\usepackage[labelfont=bf,tableposition=top]{caption} 	
\usepackage{subfig}									
\usepackage{epstopdf}								
\usepackage{rotating}
\usepackage{floatrow}
\newfloatcommand{capbtabbox}{table}[][\FBwidth]
\newsavebox{\bigleftbox}
\floatstyle{plaintop}
\restylefloat{table}
\usepackage[table,xcdraw]{xcolor}
\usepackage{soul}
\usepackage{mathrsfs}  
\usepackage{multirow}
\usepackage{amsmath}
\usepackage{amssymb}
\usepackage[capitalise]{cleveref}
\crefformat{appendix}{#2#1#3}
\crefname{rmk}{Remark}{Remarks}
\usepackage{titlesec}
\usepackage{etoolbox}
\usepackage{lipsum}
\usepackage{chngpage}
\usepackage{wrapfig}
\usepackage{subfiles}
\usepackage{geometry}

\usepackage{tikz-cd}

\newtheorem{remark}{Remark}[section]

\renewcommand{\vec}[1]{\boldsymbol{#1}}

\renewcommand{\u}{\vec{u}}

\newcommand{\E}{\mathcal{E}}
\newcommand{\Sent}{\mathcal{S}}

\newcommand{\tfin}{t_{\textsc{fin}}}

\newcommand{\Dom}{\Omega}

\makeatletter
\def\H{\@ifnextchar\bgroup\H@arg\H@noarg}
\newcommand{\H@arg}[1]{H^1( #1 )}
\newcommand{\H@noarg}{H^1}
\def\Ho{\@ifnextchar\bgroup\Ho@arg\Ho@noarg}
\newcommand{\Ho@arg}[1]{H^1_0( #1 )}
\newcommand{\Ho@noarg}{H^1_0}
\def\BH{\@ifnextchar\bgroup\BH@arg\BH@noarg}
\newcommand{\BH@arg}[1]{\vec{H}^1( #1 )}
\newcommand{\BH@noarg}{\vec{H}^1}
\def\BHo{\@ifnextchar\bgroup\BHo@arg\BHo@noarg}
\newcommand{\BHo@arg}[1]{\vec{H}^1_{\vec{0}}( #1 )}
\newcommand{\BHo@noarg}{\vec{H}^1_{\vec{0}}}
\def\Hdiv{\@ifnextchar\bgroup\Hdiv@arg\Hdiv@noarg}
\newcommand{\Hdiv@arg}[1]{\vec{H}( \text{div}; #1 )}
\newcommand{\Hdiv@noarg}{\vec{H}(\text{div})}
\def\Hdivo{\@ifnextchar\bgroup\Hdivo@arg\Hdivo@noarg}
\newcommand{\Hdivo@arg}[1]{\vec{H}_{\vec{0}}( \text{div}; #1 )}
\newcommand{\Hdivo@noarg}{\vec{H}_{\vec{0}}(\text{div})}
\def\L{\@ifnextchar\bgroup\L@arg\L@noarg}
\newcommand{\L@arg}[1]{L^2( #1 )}
\newcommand{\L@noarg}{L^2}
\def\Lp{\@ifnextchar\bgroup\Lp@arg\Lp@noarg}
\newcommand{\Lp@arg}[1]{L^p( #1 )}
\newcommand{\Lp@noarg}{L^p}
\def\Cz{\@ifnextchar\bgroup\Cz@arg\Cz@noarg}
\newcommand{\Cz@arg}[1]{\mathcal{C}^0( #1 )}
\newcommand{\Cz@noarg}{\mathcal{C}^0}
\def\Cinf{\@ifnextchar\bgroup\Cinf@arg\Cinf@noarg}
\newcommand{\Cinf@arg}[1]{\mathcal{C}^\infty( #1 )}
\newcommand{\Cinf@noarg}{\mathcal{C}^\infty}
\def\Cinfo{\@ifnextchar\bgroup\Cinfo@arg\Cinfo@noarg}
\newcommand{\Cinfo@arg}[1]{\mathcal{C}^\infty_0( #1 )}
\newcommand{\Cinfo@noarg}{\mathcal{C}^\infty_0}
\def\Cneg{\@ifnextchar\bgroup\Cneg@arg\Cneg@noarg}
\newcommand{\Cneg@arg}[1]{\mathcal{C}^{-1}( #1 )}
\newcommand{\Cneg@noarg}{\mathcal{C}^{-1}}
\def\CI{\@ifnextchar\bgroup\CI@arg\CI@noarg}
\newcommand{\CI@arg}[1]{\mathcal{C}^{1}( #1 )}
\newcommand{\CI@noarg}{\mathcal{C}^{1}}
\def\CII{\@ifnextchar\bgroup\CII@arg\CII@noarg}
\newcommand{\CII@arg}[1]{\mathcal{C}^{2}( #1 )}
\newcommand{\CII@noarg}{\mathcal{C}^{2}}
\makeatother

\newcommand{\IR}{\mathbb{R}}

\makeatletter
\def\Proj{\@ifnextchar\bgroup\Proj@arg\Proj@noarg}
\newcommand{\Proj@arg}[1]{\mathscr{P}_{#1}}
\newcommand{\Proj@noarg}{\mathscr{P}}
\makeatother

\newcommand{\jump}[1]{\mbox{$[\![ #1 ]\!]_{1,2}^{i}$}}
\newcommand{\avg}[1]{\mbox{$\{\!\>\!\>\!\!\!\{ #1 \}\!\>\!\>\!\!\!\}_{1,2}^{i}$}}

\renewcommand{\d}[1]{\,\mathrm{d}#1}
\newcommand{\dDom}{\d{V}}

\newcommand{\dbdy}{\d{S}}
\newcommand{\dd}[2]{\frac{\mathrm{d} #1 }{\mathrm{d} #2}}
\newcommand{\DDt}[1]{\frac{\mathrm{D} #1 }{\mathrm{D} t}}

\newcommand{\pdd}[2]{\frac{\partial #1 }{\partial #2}}

\renewcommand{\Re}{\mathbb{R}}

\title{Thermodynamics of surfactant-enriched binary-fluid systems}

\author{Tom~B.~van~Sluijs\corref{cor}}
\ead{t.b.v.sluijs@tue.nl}
\author{Stein~K.F.~Stoter}
\ead{k.f.s.stoter@tue.nl}
\author{E.~Harald~van~Brummelen}
\ead{e.h.v.brummelen@tue.nl}

\cortext[cor]{Corresponding author: T.B.~van~Sluijs ({t.b.v.sluijs@tue.nl})}
\address{Eindhoven University of Technology, PO Box 513, 5600 MB Eindhoven, The Netherlands}

\date{\today}

\begin{document}

\begin{abstract}
    Surface-active agents (surfactants) release potential energy as they migrate from one of two adjacent fluids onto their fluid-fluid interface, a process that profoundly impacts the system's energy and entropy householding. 
    The continuum thermodynamics underlying such a surfactant-enriched binary-fluid system has not yet been explored comprehensively. 
    In this article, we present a mathematical description of such a system, in terms of balance laws, equations of state, and permissible constitutive relations and interface conditions, that satisfies the first and second law of thermodynamics. The interface conditions and permissible constitutive relations are revealed through a Coleman-Noll analysis. We characterize the system's equilibrium by defining equilibrium equivalences and study an example system. With our work, we aim to provide a systematically derived framework that combines and links various elements of existing literature, and that can serve as a thermodynamically consistent foundation for the (numerical) modeling of full surfactant-enriched binary-fluid systems.
\end{abstract}


\maketitle

\newpage
\section{Introduction}

Surface active agents (surfactants) are amphiphilic molecules, meaning that they comprise both hydrophobic and hydrophilic parts, that minimize potential energy by migrating to fluid-fluid interfaces to locate these hydrophobic and hydrophilic parts in the respective phases. On interfaces, these molecules reduce the surface tension due to a reduction of the interfacial energy density by disrupting molecular bonds \cite{Manikantan2020,Jaensson2021}. They are of key importance in numerous and widely varied settings, ranging from the coffee-stain effect \cite{Deegan1997,Lohse2021,Tan2023} and household detergents \cite{Hellmuth2016502} to industrial applications such as emulsifiers in the food industry \cite{SAGIS201459,PIAZZA2008420,C3SM51770E,DAN2013302,JANSSEN2020105771}, inkjet printing \cite{Wijshoff2017,Gaalen2021149,Gaalen2022892,Antonopoulou2021,Gaalen2021622,Venditti2022,Lohse2021} and biomedical applications \cite{D0SM00415D,RUHS2014174,D0SM01232G,WU2012464,HOLLENBECK20142245,Hermans20158048,Tein2020}.

Given the wide range of applications, understanding the complex influence of surfactants on a binary fluid system is crucial. Models are indispensable for understanding and predicting the impact of surfactants on dynamical processes and, consequently, many types of models exist. The dynamics of interfaces have been explored for material and immaterial interfaces~\cite{Cermelli2005}, in conjunction with inviscid as well as viscous (e.g., Boussinesq-Scriven) interface stress responses \cite{SCRIVEN1960}. The surfactant transport from, to, and across the interface can be modeled based on statistical mechanics considerations, \cite{Chang1995}. Various so-called ``isotherms'' have been formulated that describe the bulk-to-interface distribution of surfactant concentration \cite{Chang1995}, for example Henry isotherm, Langmuir isotherm, Freundlich isotherm, Volmer isotherm, and Frumkin isotherm \cite{Aveyard1973,Chang1995,Garcke2013}. Implicitly, these isotherms also define the surfactant-concentration influence on the surface tension \cite{Aveyard1973,Chang1995}, which dictates the Young-Laplace pressure across the interface, as well as the Marangoni flow along the interface~\cite{Gaalen2020}. 

In order to (computationally) describe a complete binary-fluid surfactant system, these component-level models must be combined. Notable efforts are those by Venerus and {\"O}ttinger who describe the general thermodynamics of multi-phase systems with interfaces (but without surfactants)~\cite{Venerus2017}, Zhang et al. who derive a continuum model with insoluble surfactants on the interface \cite{Zhang2014}, and works by Garcke, Thiele, Zhao and co-workers that introduce soluble 
surfactants~\cite{Garcke2013,Thiele2016,Zhao2021}. Such system-level models enable numerical study of complex cases, as overviewed in \cite{Lohse2021}. Examples range from the investigation of the influence of surfactants on topological changes \cite{Kamat2018,Kamat2020} to studies on gravitational effects in surfactant-enriched multi-phase flows \cite{Li2019a,Li2019b} as well as the effect of surfactants on evaporation \cite{Tan2016,Diddens2017}. 

These continuum models are based on systems of partial differential equations equipped with suitable auxiliary conditions; state variables characterize the state of physical systems and the dynamic evolution of these states is governed by balance-laws. The balance laws involve auxiliary variables that drive the evolution, for which closure relations follow from constitutive relations that describe the material response to the system's state based on physical parameters. The laws of thermodynamics dictate which system behavior is permitted. Coleman and Noll introduce the application of a thermodynamic framework to a classic continuum-mechanics system~\cite{Coleman1963}. In this setting, the first and second law of thermodynamics, describing the balance of total energy and entropy, respectively, lead to general conditions for admissible constitutive relations. 


Despite the importance of binary-fluid-surfactant systems and the wide range of existing work, the available literature lacks a complete and coherently derived continuum-thermodynamics-based model. In this article, we present such a systematic derivation, where we methodically retrieve elements of the theory as presented by Venerus and {\"O}ttinger~\cite{Venerus2017}, and Garcke, Lam, and Stinner~\cite{Garcke2013}, and found in other earlier cited works. Specifically, we consider a general binary-fluid-surfactant system, under the (standard) assumption that the mass carried by the interface is negligible. We use a procedure and argumentation analogous to Coleman and Noll~\cite{Coleman1963} to derive thermodynamically imposed constraints on constitutive relations, equations of state, and interface-bulk coupling conditions. 
In particular, our derivation naturally yields an equation of state for the interface pressure, distinctly highlighting energetic and entropic contributions. Similarly, we obtain the Young-Laplace pressure jump over an interface, and the thermal and concentration-based Marangoni stresses from the interface momentum balance. Additionally, we acquire an expression for surfactant adsorption and an equilibrium surfactant concentration distribution between the bulk fluids and the interface. This expression balances the thermal entropy increase via the release of mixture energy and the effect of configurational entropy. Finally, we showcase the non-triviality of the definition of equilibrium, derive the equilibrium equivalences for a binary-fluid surfactant system, and end by giving an example of equilibrium states as a function of temperature.

The remainder of this article is structured as follows: in \cref{sec:systemdescription} we describe the physical system in terms of the relevant state variables and introduce the modeling paradigm. The system's evolution equations are derived in \cref{sec:balance_laws}, based on first principles, and in \cref{sec:Coleman_Noll} we perform a Coleman-Noll analysis to derive the permissible constitutive relations, interface conditions, and necessary equations of state. In \cref{sec:special_cases} we consider the equilibrium equivalences of the system and show an equilibrium example. Finally, we draw conclusions in \cref{sec:conclusions}.


\newpage
\section{System description}\label{sec:systemdescription}
This section presents the considered binary-fluid-surfactant system and modeling paradigm. The binary-fluid system is set on an ambient domain, and we derive the balance laws for an arbitrary control volume in this ambient domain. In addition, we introduce the state variables, auxiliary variables, and state functions, which are used in the balance laws. The modeling paradigm consists of postulates and modeling assumptions.

\subsection{Setting and notation}
\label{sec:SetNot}
\begin{figure}
    \centering
    \includegraphics[width=\textwidth]{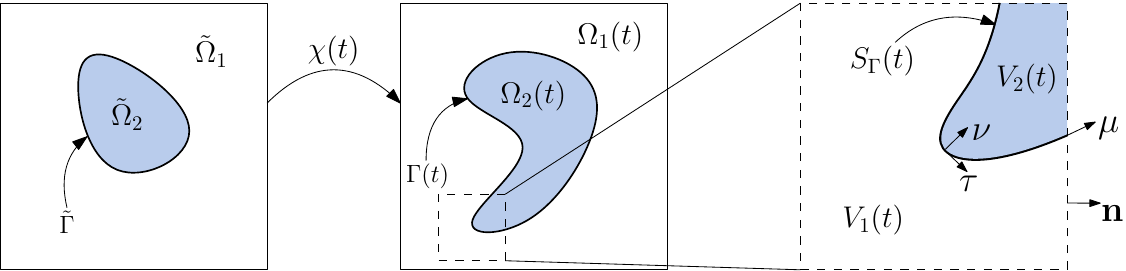}
    \caption{Schematic depiction of the modeled system, illustrated by the reference domain, domain, and control volume, from left to right respectively.}
    \label{fig:domain_map}
\end{figure}

We consider a binary-fluid-surfactant system composed of two immiscible fluid species separated by an interface, with a surfactant that is soluble in both fluid species. To provide a setting for the binary-fluid-surfactant system, we regard a time interval $(0,\tfin)\subseteq\IR_{>0}$ and a spatial domain $\Dom\subset\IR^d$ $(d=2,3)$. The two fluid species reside in two complementary bulk domains $\Omega_1(t)$ and $\Omega_2(t)$, and are separated by an interface $\Gamma(t)=\partial\Omega_1\cap\partial\Omega_2$ of co-dimension~1; see~\cref{fig:domain_map} for an illustration. We assume that the time-dependent domains $\Dom_i(t)$ are images of initial reference configurations~$\tilde{\Dom}_i$ under a time-dependent bijective deformation map $\vec{\chi}(t,\cdot):\Dom\to\Dom$ corresponding to the flow of the binary fluid. That is,
$\Dom_i(t)=\vec{\chi}(t;\tilde{\Dom}_i)$, and~$\vec{\chi}(0,\cdot)=\vec{I}$ and $\tilde{\vec{u}}=\partial_t\vec{\chi}$ corresponds to the velocity of the binary fluid, expressed in the reference configuration. The binary-fluid velocity in the actual time-dependent configuration is $\vec{u}(t,\cdot)=\tilde{\vec{u}}(t,\vec{\chi}^{-1}(t,\cdot))$. We denote the restriction $\vec{\chi}|_{\Dom_i}$ by $\vec{\chi}_i$ and extend this notational convention to other functions defined on~$\Dom$.
As we will allow the binary-fluid components to slip along the interface, $\vec{\chi}$ is not generally continuous across the interface in its tangential component. We assume that $\vec{\chi}_i$ and its inverse are continuous, and that $\vec{\chi}$ is continuous across the interface in its normal component. The latter implies that we will not regard topological changes of the interface.

To derive the governing equations for the binary-fluid-surfactant system, we consider a control volume $V(t)\subseteq\Dom$ that contains the bulk fluids in the subvolumes $V_i(t)=\Dom_i(t)\cap{}V(t)$ and the interface in the control surface $S(t)=\Gamma(t)\cap{}V(t)$. For any point on the interface, we denote by $\boldsymbol{\nu}$ the unit normal vector on the interface, directed external to~$\Dom_1(t)$, and by $\boldsymbol{\tau}$ the tangent plane to~$\Gamma$. For any point on the intersection of the interface with the control-volume boundary,~$\Gamma(t)\cap\partial{}V(t)$, we denote by~$\boldsymbol{\mu}$ the unit co-normal vector; see~\cref{fig:domain_map}. The external unit vector to $\partial{}V(t)$ is indicated by~$\boldsymbol{n}$.

In the model that we consider, the state of the binary-fluid-surfactant system is characterized by the state $q$ which contains the fields of state variables, as shown in \cref{tab:variables}.
\begin{table}
\caption{Variables, with $i\in\{1,2,\Gamma\}$ for bulk 1, bulk 2 and interface quantities.} \label{tab:variables}
\begin{tabular}{l|c}
State variables          & Symbol            \\ \hline 
Velocity                 & $\vec{u}_i  $     \\ 
Pressure                 & $p_i $            \\ 
Surfactant concentration & $c_i $            \\ 
Temperature              & $\theta_i$        \\ 
Domain map               & $\vec{\chi}_i$          \\ 
\multicolumn{2}{c}{}\\
Auxiliary variables      & Symbol            \\ \hline 
Deviatoric stress        & $\vec{\tau}_{i} $ \\ 
Concentration flux       & $\vec{J}_i $      \\ 
Heat flux                & $\vec{q}_i  $     \\ 
Heat source              & $\Sigma_i $       \\ 
\multicolumn{2}{c}{}\\
State functions          & Symbol            \\ \hline 
Mixture energy density   & $G_i( c_i ) $     \\ 
Entropy density          & $s_{i}(e_{\rm{th},i}, c_i ) $ \\ 
Thermal energy density   & $e_{\rm{th},i}(\theta_i)   $  \\ 
\end{tabular}
\end{table}

\subsection{Postulates and model assumptions}\label{subsec:postulatesassumptions}
We model the binary-fluid-surfactant system in the framework of continuum thermodynamics. The general postulates of this framework comprise conservation laws and the laws of thermodynamics. Specifically, the balance of mass, momentum, concentration, and thermal energy. The first law of thermodynamics pertains to conservation of energy. We consider three contributions to the total energy: kinetic energy, thermal energy, and mixture energy. The second law of thermodynamics dictates non-negative entropy production. Within this framework, irreversible processes produce entropy and reversible processes do not. 

Additionally, we define the following \textit{model assumptions} which will be used throughout the remainder of this article:
\begin{enumerate}[label=\mbox{MA-\arabic*},leftmargin=3em]\setlength\itemsep{-3pt}
    \item The bulk phases have constant mass density. \label{MA1}
    \item The interface neither carries nor transfers mass.\label{MA2}
    \item There are no exogenous sources to any of the balance laws. \label{MA3}
\end{enumerate}
Assumption~\ref{MA1} implies that the presence of surfactants does not significantly affect the mass density in the bulk domains. This assumption is in agreement with the fact that in most practical applications of surfactants, the surfactants are added in relatively low concentrations and, hence, their overall contribution to mass density is small. Assumption~\ref{MA2} is a standard assumption in most models of binary-fluid flows. Because the thickness of a binary-fluid interface is generally only a few~\r{A}~\cite{Braslau:1988bu,Townsend:1991bs}, the mass density with respect to surface measure is negligible. For instance, the kinetic energy of the interface is orders of magnitude smaller than its surface tension. Conversely, if surface mass density is not disregarded, then the equation of motion for the interface must be retained in the model, which yields a profound complication; see, for instance,~\cite{Scriven:1960}. The second negation in~\ref{MA2} means that there is no transfer of mass between the bulk phases. It is to be noted that~\ref{MA2} does not preclude the transport of surfactants from the bulk phases to the interface and vice versa. Even if surfactant is present at the interface in large concentrations (with respect to surface measure), the surface mass density is generally negligible nonetheless. Assumption~\ref{MA3} is non-essential, and merely serves to facilitate the presentation.

The behavior of the binary-fluid-surfactant system is to a large extent determined by properties of the state functions in Table~\ref{tab:variables}. The state functions are point-wise functionals of the state variables. We insist that the state functions  satisfy the following \textit{state-function assumptions}:

\begin{enumerate}[label=\mbox{SA-\arabic*},leftmargin=2.7em]\setlength\itemsep{-3pt}
\item The mixture energies~$G_i: \Re_{\geq 0}\rightarrow \Re$ are twice continuously differentiable and strictly convex. \label{SA1}
\item The entropy density $s_i: (\Re_{\geq 0},\Re_{\geq 0})\rightarrow \Re$ is twice continuously differentiable and strictly concave, and is strictly increasing with respect to its first argument (viz. the thermal energy density). \label{SA2}
\item The thermal-energy density $e_{\rm{th},i}: \Re_{\geq 0}\rightarrow\Re_{\geq 0}$ is a strictly increasing function. \label{SA3}
\end{enumerate}
To further elucidate the model, the arguments of $G_i$, $s_i$ 
and~$e_{\rm{th},i}$ have been indicated in Table~\ref{tab:variables}.

\newpage
\section{Conservation laws}
\label{sec:balance_laws}

A full description governing the evolution of variables in~\cref{tab:variables} will require balance laws, constitutive relations, and equations of state. In this section, we focus on the balance laws following from the conservation of mass, momentum, and surfactant concentration, as well as the conservation of thermal energy and total energy, i.e., the first law of thermodynamics.

\subsection{Balance laws in bulk phases}

The transport of mass, momentum, surfactant concentration, and thermal energy can, at the continuum level, be described by conservation laws that involve fluxes, sources, and sinks. The partial differential equations then follow from balance equations at arbitrarily chosen control volumes. Despite the textbook nature of this material, we present it here to facilitate its adaptation on sub-manifolds, i.e., the interface, later on.

Consider an arbitrary extensive quantity $C$, the total of which is defined on a material control volume $V(t)$, i.e., one that moves with velocity~$\vec{u}$, as the integral of its respective density~$c$:
\begin{align}
    C_V(t) = \int\limits_{V(t)} c \d{V}
\end{align}
The time-rate of change of $C_V$ depends on the flux~$\vec{J}$ through the boundary of the control volume and on the source density~$r$ in the control volume:
\begin{align}
    \dd{C_V}{t} = \dd{}{t}\int\limits_{V(t)} c \d{V} = - \int\limits_{\partial V(t)} \vec{J}\cdot \vec{n} \d{S} + \int\limits_{V(t)} r \d{V} \,.
\end{align}
Applying Reynolds' transport theorem and the divergence theorem yields:
\begin{align}
    \int\limits_{V(t)} \pdd{c}{t} \d{V} + \int\limits_{\partial V(t)} c\,\vec{u}\cdot\vec{n} \d{S} = \int\limits_{V(t)} \pdd{c}{t} \d{V} + \int\limits_{V(t)} \nabla\cdot( c\vec{u} ) \d{V} = - \int\limits_{V(t)} \nabla\cdot\vec{J} \d{V} + \int\limits_{V(t)} r \d{V} \,.
\end{align}
As this equality holds for arbitrary choice of $V(t)$, the integrands must coincide:
\begin{align}
    \DDt{c} + c \, \nabla\cdot \vec{u}  = - \nabla\cdot\vec{J} + r \,. \label{cons_general_bulk}
\end{align}
with $\DDt{} := \pdd{}{t} + \vec{u}\cdot\nabla$ the total derivative.

Specifying the general balance law~\eqref{cons_general_bulk} to the balance of mass, with density $\rho_i$, for which there is no driving flux and there are no sources, we obtain:
\begin{align}
   \DDt{\rho_{i}} + \rho_{i} \nabla\cdot \vec{u}_{i} &=0 \,, \label{masscons1}
\end{align}
From assumptions~\ref{MA1} and~\ref{MA3} in~\cref{subsec:postulatesassumptions}, it follows that the material derivative of the mass density is zero. Equation~\eqref{masscons1} the reduces to the condition that the velocity field is solenoidal:
\begin{align}
   \nabla\cdot \vec{u}_{i}=0 \,. \label{cons_mass_bulk}
\end{align}
Solenoidality of~$\vec{u}$ implies that the second term in the left member of the general balance equation~\eqref{cons_general_bulk} vanishes. The balance of linear momentum, corresponding to density $\rho_{i}\vec{u}_{i}$ and with the (negative of the) Cauchy stress $\vec{\sigma} = \vec{\tau}_{i}-p_{i} \vec{I}$ as driving flux,
the surfactant concentration corresponding to density $c^i$ and flux $\vec{J}_{i}$, and the thermal energy with density $ e_{\rm{th},i} $, flux $\vec{q}_{i}$ and internal source $\Sigma_{i}$, can be condensed into, respectively:
\begin{align}
    \DDt{(\rho_{i}\vec{u}_{i}) } &= - \nabla p_{i} + \nabla\cdot\vec{\tau}_{i}  \,, \label{cons_momentum_bulk} \\
   \DDt{c_{i}}  &= - \nabla\cdot\vec{J}_{i} \,, \label{cons_concentration_bulk} \\
    \DDt{ e_{\rm{th},i}} &= - \nabla\cdot\vec{q}_{i} + \Sigma_{i} \,. \label{cons_thermalenergy_bulk}
\end{align}

\subsection{Interface balance laws}
Next, we consider a material control surface $S(t)\subset\Gamma(t)$ that resides on the interface, i.e., a sub-manifold moving through the higher-dimensional volumetric domain. The generic balance equation for such a control surface reads:
\begin{align}
    \dd{}{t} \int\limits_{S(t)} c_{\Gamma} \d{S} = - \int\limits_{\partial S(t)} \vec{J}_{\Gamma}\cdot\vec{\mu} \d{L} + \int\limits_{S(t)} \jump{c_i (\vec{u}_i-\vec{u}_\Gamma)+\vec{J}_i} \cdot \vec{\nu} \d{S}  + \int\limits_{S(t)} r_\Gamma \d{S} \,, \label{surfbalance}
\end{align}
where $c_{\Gamma}$ and $r_\Gamma$ represent surface densities. The vector $\vec{\nu}$ is the out-of-plane normal to $S(t)$ external to~$\Omega_1$, and the vector $\vec{\mu}$ is the in-plane outward facing normal to the boundary of the surface control volume; see~\cref{fig:domain_map}. The flux~$\vec{J}_{\Gamma}$ is tangential to the interface. The jump operator $\jump{\cdot}$ in the generic surface balance law~\eqref{surfbalance} is defined as $\jump{f_i} = f_1 - f_2$. This jump term represents the flux out of the adjacent bulk domains onto the interface due to the interface material control surface moving through the volumetric domain with velocity $\vec{u}_\Gamma$.

In this setting, Reynolds' transport theorem and the divergence theorem take on the following forms:
\begin{align}
    \dd{}{t} \int\limits_{S(t)} c_{\Gamma} \d{S} &= \int\limits_{S(t)} \pdd{c_{\Gamma}}{t} \d{S} - \int\limits_{S(t)} \kappa \, c_{\Gamma} \vec{u}_\Gamma \cdot \vec{\nu} \d{S} + \int\limits_{\partial S(t)} c_{\Gamma} \vec{u}_{\Gamma}\cdot \vec{\mu} \d{L} \,, \label{surfReyn}\\
    \int\limits_{S(t)} \nabla_\Gamma \cdot \vec{v} \d{S} &= \int\limits_{\partial S(t)} \!\! \vec{v}\cdot\vec{\mu} \d{L} - \int\limits_{S(t)} \kappa \, \vec{v} \cdot \vec{\nu} \d{S} \label{surfdivthm}
\end{align}
with $\nabla_\Gamma = \vec{I}_\Gamma \cdot \nabla = (\vec{I}-\vec{\nu}\otimes\vec{\nu})\cdot \nabla$ as the surface gradient and $\kappa = -\nabla_\Gamma \cdot \vec{\nu} $ the total (additive) curvature~\cite{Bansch2001}. 
Noting that any vector~$\vec{v}$ on the interface can be decomposed into its in-plane component~$\vec{v}_{\vec{\tau}}=\mathbf{P}_{\Gamma}\vec{v}:=\vec{\nu}\times\vec{v}\times\vec{\nu}$ and out-of-plane 
component~$v_{\vec{\nu}} \vec{\nu}=(\vec{\nu}\cdot\vec{v})\vec{\nu}$ according to $\vec{v}_\Gamma = \vec{v}_{\vec{\tau}} + v_{\vec{\nu}} \, \vec{\nu}$, the surface divergence term in~\eqref{surfdivthm} can be recast into:
\begin{align}
    \nabla_\Gamma \cdot \vec{v} =  \nabla_\Gamma \cdot  \vec{v}_{\vec{\tau}} - v_{\vec{\nu}} \kappa \label{sufdifid}
\end{align}
Making use of the three identities~\eqref{surfReyn}--\eqref{sufdifid} and noting that the flux vector $\vec{J}_{\Gamma}$ is in-plane, i.e., $\vec{J}_{\Gamma}\cdot\vec{\nu} = 0$, Equation~\eqref{surfbalance} is equivalent to:
\begin{equation}
\label{eq:gensurfbal}
    \int\limits_{S(t)} \Big\{ \pdd{c_{\Gamma}}{t}  + \nabla_\Gamma\cdot ( c_{\Gamma} \vec{u}_{\vec{\tau}} ) - \kappa \, c_{\Gamma} u_{\vec{\nu}} \Big\} \d{S} 
    = \int\limits_{S(t)} \Big\{ - \nabla_\Gamma \cdot \vec{J}_{\Gamma} + \jump{c_i (\vec{u}_i-\vec{u}_\Gamma) + \vec{J}_i} \cdot \vec{\nu} + r_\Gamma \Big\}\d{S} \,.
\end{equation}
As the choice of~$S(t)$ is arbitrary, the equality in~\eqref{eq:gensurfbal} must in fact hold for the integrands, which leads to the general form of the balance law in partial-differential form:
\begin{align}
    \pdd{c_{\Gamma}}{t} + \nabla_\Gamma \cdot ( c_{\Gamma} \vec{u}_{\vec{\tau}} ) -  \kappa c_{\Gamma} u_{\vec{\nu}}  = - \nabla_\Gamma \cdot \vec{J}_{\Gamma} + \jump{c_i (\vec{u}_i-\vec{u}_\Gamma)+\vec{J}_i} \cdot \vec{\nu} + r_\Gamma\,. \label{cons_general_interface}
\end{align}

Specializing~\eqref{cons_general_interface} to conservation of mass, we obtain the following differential equation:
\begin{align}\label{eq:surfacegeneralmassbalance}
    \pdd{\rho_{\Gamma}}{t} + \nabla_\Gamma\cdot ( \rho_{\Gamma} \vec{u}_{\vec{\tau}} ) -  \kappa  \rho_{\Gamma} u_{\vec{\nu}}  = \jump{\rho_i  (\vec{u}_i-\vec{u}_\Gamma)} \cdot \vec{\nu} \,.
\end{align}

Assumption~\ref{MA2} in~\cref{subsec:postulatesassumptions} implies that the surface mass density in the left member of~\eqref{eq:surfacegeneralmassbalance} vanishes according to~$\rho_{\Gamma}\to+0$. From the second negation in~\ref{MA2} it follows that $\rho_i(\vec{u}_i-\vec{u}_{\Gamma})\cdot\vec{\nu}=0$. Accordingly, the right member of~\eqref{eq:surfacegeneralmassbalance} vanishes as well. Assumption~\ref{MA2} is therefore consistent with conservation of mass on the interface. For later reference, we notice that the second negation in~\ref{MA2} is equivalent to
\begin{equation}
\vec{u}_1\cdot\vec{\nu} =  \vec{u}_\Gamma\cdot\vec{\nu} =    \vec{u}_2\cdot\vec{\nu}\,. \label{cons_mass_surf}
\end{equation}
As a consequence, the deformation maps $\vec{\chi}_i$ introduced in~\cref{sec:systemdescription} are continuous in their normal component across the interface and, under the standing assumptions on smoothness of $\vec{\chi}_i$ and its inverse, the deformation map $\vec{\chi}:\Dom\to\Dom$ is bijective. It is to be noted that~$\vec{\chi}$ is not generally continuous across the interface in its tangential component.


Just like for the bulk phases, the mass-balance identities simplify the general conservation equation \eqref{cons_general_interface}. For the conservation of momentum, surfactant concentration, and thermal energy we then find:
\begin{align}
    \pdd{(\rho_{\Gamma} \vec{u}_\Gamma )}{t} + \nabla_\Gamma\cdot ( \rho_{\Gamma} \vec{u}_{\vec{\tau}} \otimes \vec{u}_\Gamma ) -  \kappa  \rho_{\Gamma} \vec{u}_\Gamma  u_{\vec{\nu}}  & = - \nabla_\Gamma \cdot ( p_\Gamma \vec{I}_\Gamma - \vec{\tau}_\Gamma ) + \jump{ p_i\vec{I} - \vec{\tau}_i } \cdot \vec{\nu}\,,\label{eq:SurfaceEoM}\\
    \pdd{c_{\Gamma}}{t} + \nabla_\Gamma\cdot ( c_{\Gamma} \vec{u}_{\vec{\tau}} ) -  \kappa   c_{\Gamma} u_{\vec{\nu}} &  = - \nabla_\Gamma \cdot \vec{J}_{\Gamma} +  \jump{\vec{J}_i} \cdot \vec{\nu} \,. \label{cons_concentration_interface} \\
    \pdd{e_{\rm{th},\Gamma}}{t} + \nabla_\Gamma\cdot ( e_{\rm{th},\Gamma} \vec{u}_{\vec{\tau}} ) -  \kappa   e_{\rm{th},\Gamma} \, u_{\vec{\nu}}  & = - \nabla_\Gamma \cdot \vec{q}_{\Gamma} +  \jump{\vec{q}_i} \cdot \vec{\nu} + \Sigma_\Gamma \,. \label{cons_thermalenergy_interface}
\end{align}
Thermal energy is conventionally defined with respect to mass measure. However, because we assume that the interface does not carry mass, we use a thermal energy with respect to surface measure for completeness of the derivation. 

Invoking again $\rho_\Gamma\to+0$ according to assumption~\ref{MA2}, the momentum balance~\eqref{eq:SurfaceEoM} yields an instantaneous stress balance relation:
\begin{equation}
     \jump{ \vec{\tau}_i - p_i\vec{I} } \cdot \vec{\nu} = \nabla_\Gamma \cdot \big( \vec{\tau}_\Gamma -  p_\Gamma \vec{I}_\Gamma \big) = \nabla_\Gamma \cdot \vec{\tau}_\Gamma - \nabla_\Gamma p_\Gamma - p_\Gamma \kappa \vec{\nu}  \,. \label{cons_momentum_interface}
\end{equation}
Equation~\eqref{cons_momentum_interface} conveys that the jump of the bulk surface tractions contains a contribution of the deviatoric interface stress tensor, a tangential Marangoni component and a normal Young--Laplace pressure jump, both related to the surface pressure. The surface pressure that has been introduced in the surface equation of motion~\eqref{eq:SurfaceEoM} can therefore be identified with the negative surface tension, i.e. $p_\Gamma = -\sigma_\Gamma$. It is important to note that assumption~\ref{MA2} and the corresponding reductions 
of~\eqref{eq:surfacegeneralmassbalance} to~\eqref{cons_mass_surf}, 
and of~\eqref{eq:SurfaceEoM} to~\eqref{cons_momentum_interface}, imply that 
closure relations are required for the interface velocity~$\vec{u}_{\Gamma}$ and surface pressure $p_\Gamma$. This will be further considered 
in~\cref{subsubsec:equationsofstate}.

\subsection{The first law of thermodynamics}
\label{ssec:firstlaw}

The first step in deriving coupling relations between the previously derived evolution equations is to consider the first law of thermodynamics: the conservation of the total energy of the binary-fluid-surfactant system. The total system energy comprises three contributions, viz. kinetic energy, thermal energy and mixture energy:
\begin{equation}\label{eq:energies}
    \E = \underbrace{\sum^{2}_{i=1} \int\limits_{\Omega_{i}(t)}\!\! \frac{1}{2} \rho_{i} |\vec{u}_{i}|^{2} \d{V}}_{\E_{\rm{kinetic}}} 
    + \underbrace{\sum^{2}_{i=1} \int\limits_{\Omega_{i}(t)}\!\! e_{\rm{th},i} \d{V} + \int\limits_{\Gamma(t)}\!\!  e_{\rm{th},\Gamma} \d{S}}_{\E_{\rm{thermal}}} 
    + \underbrace{\sum^{2}_{i=1} \int\limits_{\Omega_{i}(t)}\!\! G_{i}( c_{i} ) \d{V} + \int\limits_{\Gamma(t)}\!\!  G_\Gamma ( c_{\Gamma} )\d{S}}_{\E_{\rm{mixture}}} \,,
\end{equation}
where $G_{i}\left(c_{i}\right)$ and $G_\Gamma \left(c_{\Gamma}\right)$ are the bulk and interface mixture energy density, respectively. These energy densities are response functions based on the local surfactant concentrations. The interface does not contribute to the kinetic energy as its density is negligible in accordance with assumption~\ref{MA2}. Let us note that the mixture energy also contains an interface-energy contribution. In view of its dependence on the interface surfactant concentration~$c_{\Gamma}$, we have classified this term has as a mixture-energy term.

We consider a material control volume $V(t)$ and consider the time rate of change of the energy in this volume. This volume may intersect the interface, with $V_1(t) = V(t)\cap\Omega_{1}(t)$, $V_2(t) = V(t)\cap\Omega_{2}(t)$ and $S_\Gamma(t) = V(t)\cap\Gamma(t)$. It then follows that:
\begin{equation}
    \dd{\E_{V}}{t} =  \sum^{2}_{i=1} \dd{}{t} \int\limits_{V_i(t)}\!\! \Big\{ \frac{1}{2} \rho_{i} |\vec{u}_i|^{2}+G_{i}(c_{i})+e_{\rm{th},i}\Big\} \d{V}+ \dd{}{t} \int\limits_{S_\Gamma(t)}\!\! \Big\{ G_\Gamma(c_{\Gamma}) + e_{\rm{th},\Gamma} \Big\}  \d{S}  \, .
\end{equation}
Upon invoking the bulk and interface Reynolds transport theorems, the bulk and interface divergence theorems, and recalling the solenoidality of the velocity field in the bulk phases, we obtain:
\begin{equation}
\begin{split}
    \dd{\E_{V}}{t} =&\, \sum^{2}_{i=1} \int\limits_{V_i(t)}\!\! \Big\{  \DDt{} \big( \frac{1}{2} \rho_{i} |\vec{u}_i|^{2}\big)+ \DDt{}\big( G_{i}(c_{i}) \big) + \DDt{} e_{\rm{th},i}\Big\} \d{V} \\
    &\,\,\,\, + \int\limits_{S_\Gamma(t)}\!\! \Big\{ \pdd{}{t}  G_\Gamma(c_{\Gamma}) + \nabla_\Gamma  G_\Gamma(c_{\Gamma}) \cdot \vec{u}_{\vec{\tau}} +  G_\Gamma(c_{\Gamma}) \nabla_\Gamma\cdot \vec{u}_{\Gamma}  \Big\}  \d{S} \\
    & \,\,\,\, + \int\limits_{S_\Gamma(t)}\!\!  \Big\{ \pdd{}{t}  e_{\rm{th},\Gamma} + \nabla_\Gamma  e_{\rm{th},\Gamma} \cdot \vec{u}_{\vec{\tau}} +  e_{\rm{th},\Gamma} \nabla_\Gamma\cdot \vec{u}_{\vec{\tau}} -  \kappa  \,  e_{\rm{th},\Gamma} u_{\vec{\nu}} \Big\}  \d{S}  \, .
\end{split} \label{dEvdt0}
\end{equation}
%
Appropriate expressions for the rates of change of the various quantities can be deduced from the earlier conservation laws. By multiplying the momentum conservation statement~\eqref{cons_momentum_bulk} by~$\vec{u}_i$, one finds an expression for the evolution of the kinetic energy:
\begin{align}
     \DDt{\rho_i \vec{u}_i} \vec{u}_i = \DDt{ }\left( \frac{1}{2}\rho_i | \vec{u}_i | ^2 \right) = -\nabla p_i \cdot \vec{u}_i + \big( \nabla\cdot \vec{\tau}_{i} \big) \cdot \vec{u}_i\,,
\end{align}
where the first equality is due to the constant density assumption~\ref{MA1}.
Applying the chain rule and making use of Equations~\eqref{cons_concentration_bulk} and \eqref{cons_concentration_interface}, we obtain evolution equations for the bulk and interface mixture energy:
\begin{align}
     &\DDt{}\big( G_{i}(c_{i}) \big) =  G_{i}'(c_{i}) \DDt{}\big( c_{i} \big) = - G_{i}'(c_{i}) \nabla\cdot \vec{J}_i \,, \\
\begin{split}
    &\pdd{}{t}  G_\Gamma(c_{\Gamma}) + \nabla_\Gamma  G_\Gamma(c_{\Gamma}) \cdot \vec{u}_{\vec{\tau}}  = G_\Gamma'(c_{\Gamma}) \left( \pdd{c_{\Gamma}}{t} + \nabla_\Gamma c_{\Gamma} \cdot \vec{u}_{\vec{\tau}} \right) \\
    & \hspace{4.8cm} = G_\Gamma'(c_{\Gamma}) \left( - c_{\Gamma}  \nabla_\Gamma\cdot \vec{u}_{\Gamma} - \nabla_\Gamma \cdot \vec{J}_{\Gamma} +  \jump{\vec{J}_i} \cdot \vec{\nu} \right)\,.
\end{split}
\end{align}
Substituting the above equalities, as well as the balance laws for thermal energy, Equations~\eqref{cons_thermalenergy_bulk} and \eqref{cons_thermalenergy_interface}, into Equation~\eqref{dEvdt0} results in:
\begin{equation}
\begin{split}
    &\dd{\E_{V}}{t} = \sum^{2}_{i=1} \int\limits_{V_i(t)}\!\! \Big\{ -\nabla p_i \cdot \vec{u}_i +\big( \nabla\cdot \vec{\tau}_{i} \big) \cdot \vec{u}_i - G_{i}'(c_{i}) \nabla\cdot \vec{J}_i -\nabla\cdot \vec{q}_i + \Sigma_i \Big\} \d{V} \\
    &\,\,\,\, + \int\limits_{S_\Gamma(t)}\!\! \Big\{  - G_\Gamma'(c_{\Gamma}) \nabla_\Gamma \cdot \vec{J}_{\Gamma} + G_\Gamma'(c_{\Gamma}) \jump{\vec{J}_i} \cdot \vec{\nu} + \big[ G_\Gamma(c_{\Gamma}) - c_{\Gamma} 
 G_\Gamma'(c_{\Gamma}) \big] \,  \nabla_\Gamma\cdot \vec{u}_{\Gamma} \Big\} \d{S} \\
    & \,\,\,\, + \int\limits_{S_\Gamma(t)}\!\!  \Big\{ - \nabla_\Gamma \cdot \vec{q}_{\Gamma} +  \jump{\vec{q}_i} \cdot \vec{\nu} + \Sigma_\Gamma \Big\}  \d{S}  \, .
\end{split} \label{dEvdt1} \raisetag{1.25cm}
\end{equation}

To distinguish between exogenous and endogenous influences, we perform integration by parts on the bulk integrals. In particular, for the first and second term in the right member of~\eqref{dEvdt1} we obtain:
\begin{multline}
\sum^{2}_{i=1} \int\limits_{V_i(t)}  
\!\! \Big\{ -\nabla p_i \cdot \vec{u}_i+\big( \nabla\cdot \vec{\tau}_{i} \big) \cdot \vec{u}_i\Big\}\d{V}
=
\sum^{2}_{i=1} \int\limits_{V_i(t)}  
\!\! \Big\{ p_i\nabla\cdot\vec{u}_i-\vec{\tau}_{i}:\nabla\vec{u}_i\Big\}\d{V}
\\
+
\sum^{2}_{i=1}\int\limits_{\partial V_{i,\rm{ext}} (t)}\!\!
\big[ \vec{\tau}_{i} \cdot\vec{n} - p_i \vec{n} \big] \cdot\vec{u}_i\d{S}
+
\int\limits_{S_\Gamma(t)}\!\!
 \avg{\vec{u}_i}\cdot\jump{ \vec{\tau}_i - p_i\vec{I} } \cdot \vec{\nu}
 +
 \jump{\vec{u}_i}\cdot\avg{ \vec{\tau}_i - p_i\vec{I} } \cdot \vec{\nu}\d{S}\,,
 \label{eq:elab01}
\end{multline}
where $\avg{(\cdot)_i}=\frac{1}{2}(\cdot)_1+\frac{1}{2}(\cdot)_2$ denotes the average across the interface, $S_\Gamma=V\cap\Gamma$ and 
$\partial V_{i,\rm{ext}}=\partial V_i\setminus S_\Gamma$. Note that $\jump{\vec{u}_i}=\jump{\vec{u}_{\vec{\tau},i}}$ by virtue of the continuity of the normal component of the velocities across the interface according to Equation~\eqref{cons_mass_surf}. In addition, $\nabla\cdot\vec{u}_i$ vanishes on account of~(\ref{cons_mass_bulk}). 
By adding the partition of zero
$\vec{u}_{\Gamma}-(\vec{u}_{\vec{\tau}}+\vec{u}_{\vec{\nu}})$ to~$\avg{\vec{u}_i}$ in~(\ref{eq:elab01}), it follows from the reduced interface-momentum balance \eqref{cons_momentum_interface} in conjunction with~\eqref{dEvdt1} that 
\begin{equation}
\begin{split}
    &\dd{\E_{V}}{t} = \sum^{2}_{i=1} \int\limits_{V_i(t)}\!\! \Big\{ \Sigma_i - \vec{\tau}_{i} :  \nabla\vec{u}_i + \nabla G_{i}'(c_{i}) \cdot \vec{J}_i \Big\} \d{V} \\
     &\,\,\,\, + \sum^{2}_{i=1} \int\limits_{\partial V_{i,\rm{ext}} (t)}\!\! \Big\{ \big[ \vec{\tau}_{i} \cdot\vec{n} - p_i \vec{n} \big] \cdot\vec{u}_i  - G_{i}'(c_{i}) \vec{J}_i \cdot \vec{n} - \vec{q}_i\cdot\vec{n} \Big\} \d{S} \\
    &\,\,\,\, + \int\limits_{S_\Gamma(t)}\!\! \Big\{\Sigma_\Gamma  - \vec{\tau}_\Gamma : \nabla_\Gamma\vec{u}_\Gamma  + p_\Gamma \nabla_\Gamma\cdot \vec{u}_{\Gamma} +\nabla_\Gamma \cdot \big( \vec{\tau}_\Gamma -  p_\Gamma \vec{I}_\Gamma \big) \cdot \left(\avg{\vec{u}_{\vec{\tau},i}}-\vec{u}_{\vec{\tau}}\right)\Big\}  \d{S} \\
     &\,\,\,\, + \int\limits_{S_\Gamma(t)}\!\! \Big\{\jump{\vec{u}_{\vec{\tau},i}}\cdot \avg{\vec{\tau}_i-p_i\vec{I}}\cdot\vec{\nu} + \nabla_\Gamma G_\Gamma'(c_{\Gamma}) \cdot \vec{J}_{\Gamma}  + \jump{\big( G_\Gamma'(c_{\Gamma}) - G'_i(c_i) \big) \vec{J}_i } \cdot \vec{\nu}   \\[-0.3cm]
    &\qquad\qquad\quad + \big[ G_\Gamma(c_{\Gamma}) - c_{\Gamma} 
 G_\Gamma'(c_{\Gamma}) \big] \,  \nabla_\Gamma\cdot \vec{u}_{\Gamma} \Big\}  \d{S} \\
    & \,\,\,\, + \int\limits_{\partial S_\Gamma(t)}\!\!  \Big\{ \big[ \vec{\tau}_{\Gamma} \cdot\vec{\mu} - p_\Gamma \vec{\mu} \big] \cdot\vec{u}_\Gamma - G_\Gamma'(c_{\Gamma}) \vec{J}_{\Gamma} \cdot \vec{\mu}- \vec{q}_{\Gamma} \cdot\vec{\mu} \Big\}  \d{L}  \, .
\end{split} \label{dEvdt}\raisetag{1.25cm}
\end{equation}
The second and last integrals represent exogenous influences on the bulk and interface: the mechanical work exerted on the boundary of the control domains, the in- or outflux of mixture energy and the in- or outflux of thermal energy. The first law of thermodynamics insists that energy is conserved up to these exogenous terms, meaning that the remaining integrals must vanish. As the control volume~$V$ is arbitrary, the integrands must vanish. It therefore holds that:
\begin{align}
\Sigma_i\, &= \vec{\tau}_{i} :  \nabla\vec{u}_i - \nabla G_{i}'(c_{i}) \cdot \vec{J}_i \,,  \label{balance_heatsource_bulk} \\
\Sigma_\Gamma &= \vec{\tau}_\Gamma : \nabla_\Gamma\vec{u}_\Gamma  - p_\Gamma  \nabla_\Gamma\cdot \vec{u}_{\Gamma} - \nabla_\Gamma G_\Gamma'(c_{\Gamma}) \cdot \vec{J}_{\Gamma}-\nabla_\Gamma \cdot \big( \vec{\tau}_\Gamma -  p_\Gamma \vec{I}_\Gamma \big) \cdot \left(\avg{\vec{u}_{i,\vec{\tau}}}-\vec{u}_{\vec{\tau}}\right)
\notag
\\ 
&\phantom{=}-\left(\avg{\vec{\tau}_i-p_i\vec{I}}\cdot\vec{\nu}\right)\cdot\jump{\vec{u}_{\vec{\tau},i}} - \jump{\big( G_\Gamma'(c_{\Gamma}) - G'_i(c_i) \big) \vec{J}_i } \cdot \vec{\nu} 
\notag
\\
&\phantom{=}
-\big[ G_\Gamma(c_{\Gamma}) - c_{\Gamma} 
 G_\Gamma'(c_{\Gamma}) \big] \,  \nabla_\Gamma\cdot \vec{u}_{\Gamma} \,. \label{balance_heatsource_interface} 
\end{align}

\newpage
\section{Constitutive relations and equations of state}
\label{sec:Coleman_Noll}
Constitutive relations for the fluxes must be such that the evolution of the system adheres to the second law of thermodynamics: any subsystem's entropy is non-decreasing up to exogeneous influences. This section exposes the permissible forms of the constitutive relations through a Coleman-Nole procedure.

\subsection{The second law of thermodynamics}\label{subsec:secondlaw}
The total entropy of the system is comprised of bulk and interface contributions:
\begin{equation}
    \Sent = \sum^{2}_{i=1} \int\limits_{\Omega_i(t)} s_{i}  \d{V} + \int\limits_{\Gamma(t)}  s_{\Gamma} \d{S} \,, 
\end{equation}
with $s_{i} = s_i(e_{\rm{th},i},c_i)$ ($i \in \{1,2,\Gamma \}$) the entropy density as a function of the thermal energy and the surfactant concentration. Typically, 
the entropy density admits an additive decomposition into thermal entropy and
configurational (or mixture) entropy according to $s_i(e_{\rm{th},i},c_i) = s_{\rm{th},i}(e_{\rm{th},i}) + s_{\rm{mix},i}(c_i)$. However, we retain the combined form for generality and conciseness.

The second law of thermodynamics insists that on an arbitrary material control volume~$V(t)$, the rate of change of the total entropy cannot subceed the transport of entropy into the control volume. In the presented modeling framework, transport of entropy into the volume can be induced by transport of thermal energy and transport of surfactant concentration, through:
\begin{align}
    &\d{s_{i}} =  \pdd{s_{i}}{e_{\rm{th},i}} \d{e_{\rm{th},i}} + \pdd{s_{i}}{c_{i}} \d{c_{i}}  =: \frac{1}{\theta_{i}} \d{e_{\rm{th},i}} + \frac{1}{\eta_{i}} \d{c_{i}} \qquad  i \in \{1,2,\Gamma \} \,,
\end{align}
where $\theta_i$ may be recognized as the thermodynamic temperature of the bulk or interface, and $\eta_i$ is a mixture-energy based equivalent.
The following Clausius--Duhem-type inequality then captures the second law on an arbitrary control volume~$V$:
\begin{align}
\begin{split}
    \dd{\Sent_V}{t} &\geq - \sum^{2}_{i=1}  \int\limits_{\partial V_{\rm{ext},i}(t)}\!\! \Big\{  \frac{\vec{q}_i}{\theta_{i}} \cdot\vec{n} +  \frac{\vec{J}_i}{\eta_{i}} \cdot\vec{n} \Big\} \d{S} - \int\limits_{\partial S_\Gamma(t)}\!\! \Big\{ \frac{\vec{q}_\Gamma}{\theta_{\Gamma}} \cdot\vec{\mu} +  \frac{\vec{J}_\Gamma}{\eta_{\Gamma}} \cdot\vec{\mu}  \Big\}\d{L}  \,. 
\end{split}\label{ClausiusDuhem}\raisetag{1.2cm}
\end{align}
where, like before, $\partial V_{\rm{ext},i} = \partial V \cap \Omega_i$, and $\partial S_\Gamma$ is the boundary of $S_\Gamma$.
Integration-by-parts on these boundary terms yields expressions on the control-volume interior:
\begin{align}
\begin{split}
\dd{\Sent_V}{t} &\geq - \sum^{2}_{i=1}  \int\limits_{V_i(t)}\!\! \Big\{  \frac{1}{\theta_{i}}\nabla\cdot \vec{q}_i - \frac{1}{\theta_{i}^2}\nabla\theta_{i}\cdot \vec{q}_i +  \frac{1}{\eta_{i}}\nabla\cdot \vec{J}_i -  \frac{1}{\eta_{i}^2} \nabla \eta_{i}\cdot \vec{J}_i \Big\} \d{V} \\
&\hspace{3cm} + \int\limits_{ S_\Gamma(t)}\!\! \Big\{ \jump{ \frac{\vec{q}_i}{\theta_{i}}} \cdot\vec{\nu} + \jump{\frac{\vec{J}_i}{\eta_{i}}} \cdot\vec{\nu} \Big\}\d{S}  \\
&\qquad - \int\limits_{ S_\Gamma(t)}\!\! \Big\{ \frac{1}{\theta_{\Gamma}}\nabla_{\Gamma}\cdot \vec{q}_\Gamma - \frac{1}{\theta_{\Gamma}^2}\nabla_{\Gamma}\theta_{\Gamma}\cdot \vec{q}_\Gamma +  \frac{1}{\eta_{\Gamma}} \nabla_{\Gamma}\cdot\vec{J}_\Gamma -  \frac{1}{\eta_{\Gamma}^2}\nabla_{\Gamma}\eta_{\Gamma}\cdot \vec{J}_\Gamma \Big\}\d{S}  \,. 
\end{split}\raisetag{1.2cm} \label{dSvdtRHS}
\end{align}

Considering next the left-hand side of Equation~\eqref{ClausiusDuhem}, it follows from the bulk and interface Reynolds' transport theorems and the chain-rule of differentiation that:
\begin{align}
\begin{split}
    \dd{\Sent_V}{t}  &= \dd{}{t} \sum^{2}_{i=1} \int\limits_{V_i(t)} s_i \d{V} + \dd{}{t} \int\limits_{S_\Gamma(t)}  s_\Gamma \d{S}  \\
    &=  \sum^{2}_{i=1} \int\limits_{V_i(t)} \DDt{s_{i}}\d{V}  + \int\limits_{S_\Gamma(t)} \Big\{ \pdd{s_{\Gamma}}{t} +\nabla_\Gamma s_{\Gamma} \cdot \vec{u}_{\vec{\tau}} + s_{\Gamma} \, \nabla_\Gamma \cdot \vec{u}_{\Gamma}  \Big\}  \d{S} \\
    &=  \sum^{2}_{i=1} \int\limits_{V_i(t)} \Big\{ \frac{1}{\theta_i} \DDt{e_{\rm{th},i}} + \frac{1}{\eta_i} \DDt{c_{i}}  \Big\} \d{V}  + \int\limits_{S_\Gamma(t)} s_{\Gamma} \, \nabla_\Gamma \cdot \vec{u}_{\Gamma} \d{S}  \\
    &\quad + \int\limits_{S_\Gamma(t)} \Big\{ \frac{1}{\theta_\Gamma } \Big( \pdd{e_{\rm{th},\Gamma}}{t} +\nabla_\Gamma e_{\rm{th},\Gamma} \cdot \vec{u}_{\vec{\tau}}\Big) +\frac{1}{\eta_\Gamma} \Big(\pdd{c_\Gamma}{t}  +\nabla_\Gamma c_{\Gamma} \cdot \vec{u}_{\vec{\tau}}\Big) \Big\}  \d{S}  \,.\label{dSvdtRT}
\end{split}
\end{align}
This form now permits substitution of the balance laws for surfactant concentration and thermal energy in the bulk and on the interface, i.e., Equations~\eqref{cons_concentration_bulk},~\eqref{cons_thermalenergy_bulk},~\eqref{cons_concentration_interface}, and~\eqref{cons_thermalenergy_interface}:
\begin{align}
\begin{split}
    \dd{\Sent_V}{t}  &=  \sum^{2}_{i=1} \int\limits_{V_i(t)} \Big\{ \frac{ - \nabla\cdot\vec{q}_{i} + \Sigma_{i} }{\theta_i} + \frac{- \nabla\cdot\vec{J}_{i}}{\eta_i} \Big\} \d{V}  + \int\limits_{S_\Gamma(t)}  s_{\Gamma} \, \nabla_\Gamma \cdot \vec{u}_{\Gamma} \d{S}  \\
    &\quad + \int\limits_{S_\Gamma(t)} \Big\{ \frac{1}{\theta_\Gamma } \Big( - e_{\rm{th},\Gamma} \nabla_\Gamma \cdot \vec{u}_{\Gamma} - \nabla_\Gamma \cdot \vec{q}_{\Gamma} +  \jump{\vec{q}_i} \cdot \vec{\nu} + \Sigma_\Gamma \Big) \Big\}  \d{S}  \\
    &\quad + \int\limits_{S_\Gamma(t)} \Big\{\frac{1}{\eta_\Gamma} \Big( - c_{\Gamma} \nabla_\Gamma \cdot \vec{u}_{\Gamma} - \nabla_\Gamma \cdot \vec{J}_{\Gamma} +  \jump{\vec{J}_i} \cdot\vec{\nu} \Big) \Big\}  \d{S} \,.
\end{split}\raisetag{1.25cm} \label{dSvdtLHS}
\end{align}
Combining Equations~\eqref{dSvdtRHS} and~\eqref{dSvdtLHS} while also substituting the expressions for the bulk and interface heat sources, $\Sigma_i$ and $\Sigma_\Gamma$ according 
to Equations~\eqref{balance_heatsource_bulk} and~\eqref{balance_heatsource_interface}, leads to the inequality:
%
\begin{align}
\label{eq:balance_entropy}
\begin{split}
    &\sum^{2}_{i=1} \int\limits_{V_i(t)} \Big\{ \frac{1}{\theta_i} \vec{\tau}_{i} :  \nabla\vec{u}_i - \frac{1}{\theta_{i}^2}\nabla\theta_{i}\cdot \vec{q}_i - \Big( \frac{1}{\theta_i} \nabla G_{i}'(c_{i})  +\frac{1}{\eta_{i}^{2}} \nabla\eta_{i}\Big)\cdot \vec{J}_i \Big\} \d{V}   \\
    & + \int\limits_{S_\Gamma(t)}\!\! \Big\{ \frac{1}{\theta_\Gamma } \vec{\tau}_\Gamma : \nabla_\Gamma\vec{u}_\Gamma  - \frac{1}{\theta_{\Gamma}^2}\nabla_{\Gamma}\theta_{\Gamma}\cdot \vec{q}_{\Gamma}  - \Big( \frac{1}{\theta_\Gamma} \nabla_{\Gamma} G_{\Gamma}'(c_{\Gamma})  +\frac{1}{\eta_{\Gamma}^2} \nabla_{\Gamma}\eta_{\Gamma}\Big) \cdot \vec{J}_{\Gamma} \Big\}  \d{S} \\
    & + \int\limits_{S_\Gamma(t)}\!\! \frac{1}{\theta_{\Gamma}}\Big\{ -\nabla_\Gamma \cdot \big( \vec{\tau}_\Gamma -  p_\Gamma \vec{I}_\Gamma \big) \cdot \big(\avg{\vec{u}_{i,\vec{\tau}}}-\vec{u}_{\vec{\tau}}\big)-\left(\avg{\vec{\tau}_i-p_i\vec{I}}\cdot\vec{\nu}\right)\cdot\jump{\vec{u}_{\vec{\tau},i}} \Big\}  \d{S} \\
    &+ \int\limits_{ S_\Gamma(t)}\!\! \Big\{ \jump{ \big( \frac{1}{\theta_\Gamma}  -  \frac{1}{\theta_{i}} \big) \vec{q}_i} \cdot\vec{\nu} +  \jump{\big( -\frac{G_\Gamma'(c_{\Gamma})}{\theta_\Gamma} + \frac{G'_i (c_i)}{\theta_\Gamma} + \frac{1}{\eta_\Gamma}  - \frac{1}{\eta_{i}} \big) \vec{J}_i} \cdot\vec{\nu} \Big\}\d{S} \\
    &  - \int\limits_{S_\Gamma(t)} \!\!\! \Big( \frac{1}{\theta_\Gamma} p_\Gamma  +\frac{1}{\theta_\Gamma}\big[ G_\Gamma(c_{\Gamma}) - c_{\Gamma} 
    G_\Gamma'(c_{\Gamma}) \big]  + \frac{1}{\theta_\Gamma } e_{\rm{th},\Gamma} + \frac{1}{\eta_\Gamma} c_{\Gamma} - s_{ \Gamma} \Big)\, \nabla_\Gamma \cdot \vec{u}_{\Gamma} \d{S}  \geq 0\,.
\end{split}\raisetag{3.7cm}
\end{align}
This describes the total entropy production within the control volume at any moment during the evolution of the system. The inequality must hold for any arbitrary control volume, including control volumes that are contained within a single bulk domain or that collapse onto the interface. Based on a localization argument, this equation thus formally splits into three inequalities: one for each bulk fluid and one for the interface.

\subsection{Response functions}
In \cref{sec:balance_laws}, the closure relations for the bulk traction jump and heat sources, according to~\eqref{cons_momentum_bulk} and
~\eqref{balance_heatsource_bulk}-\eqref{balance_heatsource_interface}, respectively, have been derived directly from the postulated conservation laws. 
Completion of the PDE model (studied in its weak form in \cref{app:WF}) still requires closure relations for the remaining auxiliary variables 
in~\cref{tab:variables}, as well as equations of state for~$\vec{u}_{\Gamma}$ and $p_{\Gamma}$, and interface conditions. 

Admissible response functions are those that ensure satisfaction of the second law of thermodynamics according to Equation~\eqref{eq:balance_entropy}. In principle, any set of response functions that collectively guarantee the inequality 
in Equation~\eqref{eq:balance_entropy} is admissible. Yet, more insight in the structure of the underlying entropy production mechanisms can be attained by postulating a limited parametric dependence of the remaining response functions. We therefore proceed by introducing the following minimal \textit{constitutive class} of parametric dependences of the remaining response functions on the state variables:
\begin{equation}\label{eq:respfuncpostulate}
\begin{split}
 &\left. p_{\Gamma} = p_{\Gamma}(c_\Gamma , \theta_\Gamma)\right. \\
 &\left. \vec{u}_{\vec{\tau}}= \vec{u}_{\vec{\tau}}(\avg{\vec{u}_{i,\vec{\tau}}})\right.\\
 &\left.\begin{split}
     \vec{\tau}_{i}&=\vec{\tau}_{i}(\nabla\vec{u}_{i})\\
     \vec{q}_{i}&=\vec{q}_{i}(\nabla\theta_{i})\\
     \vec{J}_{i}&=\vec{J}_{i}(\nabla c_{i})\\
\end{split}\right\}\; i=1,2,\Gamma \,, \\
&\left.\begin{split}
     \vec{q}_{i}\big|_{\Gamma}&=\vec{q}_{i}\big|_{\Gamma}(\theta_{i},\theta_{\Gamma})\\
     \vec{J}_{i}\big|_{\Gamma}&=\vec{J}_{i}\big|_{\Gamma}(c_{i},c_{\Gamma},\theta_{i},\theta_{\Gamma})\\
\end{split}\right\}\; i=1,2 \,,
\end{split}
\end{equation}
where the gradient operator corresponds to the surface gradient in the case of $i=\Gamma$. The constitutive class~\eqref{eq:respfuncpostulate} is minimal in the sense that under further reduction of the dependences, inequality~\eqref{eq:balance_entropy} is not generally satisfied or only satisfied for trivial (i.e. vanishing) response functions.
In the subsequent subsections, we derive explicit expressions for the response functions in~\eqref{eq:respfuncpostulate} subject to~\eqref{eq:balance_entropy}.


\subsubsection{Equations of state}\label{subsubsec:equationsofstate}

Consider a system state of uniform and equaling velocity and temperature fields over all domains ($\nabla\vec{u}_i=\vec{0}$,  $\jump{\vec{u}_i} = \vec{0}$, $\nabla \theta_i=\vec{0}$, and $\frac{1}{\theta_\Gamma}-\frac{1}{\theta_i} = 0$ for $i=1,2,\Gamma$), uniform surfactant concentrations ($\nabla G_i'(c_i) = \vec{0}$ for $i=1,2,\Gamma$), and the bulk surfactant concentrations that relate to the interface concentration such that they satisfy $G'_i (c_i) - \frac{\theta_{\Gamma}}{\eta_{i}} = G_\Gamma'(c_{\Gamma})  - \frac{\theta_{\Gamma}}{\eta_\Gamma}$ (see \cref{rem:bulksurfacecons}). For such a system state, Equation~\eqref{eq:balance_entropy} reduces to:
\begin{align}
\begin{split}\label{eq:balance_entropy_stateeqs}
    & \int\limits_{S_\Gamma(t)}\!\! \Big\{  -\frac{1}{\theta_{\Gamma}}\nabla_\Gamma \cdot \big( \vec{\tau}_\Gamma -  p_\Gamma \vec{I}_\Gamma \big) \cdot \big(\avg{\vec{u}_{i,\vec{\tau}}} -\vec{u}_{\vec{\tau}}\big) \\[-0.15cm]
    & \qquad -  \Big( \frac{1}{\theta_\Gamma} p_\Gamma  +\frac{1}{\theta_\Gamma}\big[ G_\Gamma(c_{\Gamma}) - c_{\Gamma} 
    G_\Gamma'(c_{\Gamma}) \big]  + \frac{1}{\theta_\Gamma } e_{\rm{th},\Gamma} + \frac{1}{\eta_\Gamma} c_{\Gamma} - s_{ \Gamma} \Big)\, \nabla_\Gamma \cdot \vec{u}_{\Gamma} \Big\} \d{S}  \geq 0\,,
\end{split}
\end{align}
which must still be satisfied for arbitrary instances of the state variables $\avg{\vec{u}_{i,\vec{\tau}}}$, $\theta_\Gamma$ and~$c_{\Gamma}$. Given that the pressure jump over the interface $\nabla_\Gamma \cdot \big( \vec{\tau}_\Gamma -  p_\Gamma \vec{I}_\Gamma \big)$ and the interface expansion or contraction $ \nabla_\Gamma \cdot \vec{u}_{\Gamma}$ can have arbitrary signs, and the postulated dependence of $\vec{u}_{\vec{\tau}}$ exclusively on $\avg{\vec{u}_{i,\vec{\tau}}}$, and of $p_\Gamma$ exclusively on $\theta_\Gamma$ and $c_{\Gamma}$, this can only hold generally if the following equations of state apply:
\begin{align}\label{eq:tangentialinterfacevelocity}
   \vec{u}_{\vec{\tau}} & = \avg{\vec{u}_{i,\vec{\tau}}} \,, \\
\begin{split}\label{eq:surfacepressure}
    p_\Gamma &= -\big[ G_\Gamma(c_{\Gamma}) - c_{\Gamma}  G_\Gamma'(c_{\Gamma}) \big] + \theta_\Gamma \Bigg[ s_{ \Gamma} - c_{\Gamma} \pdd{s_{ \Gamma}}{c_{\Gamma}} \Bigg] -  e_{\rm{th},\Gamma} \\
    & = -\sigma_{\scriptscriptstyle G} + \theta_\Gamma \sigma_s -  e_{\rm{th},\Gamma}\,,
\end{split}
\end{align}
whereby the entropy production of \eqref{eq:balance_entropy_stateeqs} equals zero. Let us note that the equations og state~\eqref{eq:tangentialinterfacevelocity} and~\eqref{eq:surfacepressure} are consequences of~\eqref{eq:balance_entropy} in the constitutive class~\eqref{eq:respfuncpostulate}. For instance, if a different constitutive class containing the response function $\tilde{p}_{\Gamma} = \tilde{p}_{\Gamma}(c_\Gamma , \theta_\Gamma, \nabla_\Gamma \cdot \vec{u}_\Gamma )$ is considered, then Equation~\eqref{eq:surfacepressure} could include a compressibility-based constitutive (dissipative) response according to~$\tilde{p}_{\Gamma} = p_\Gamma - K \nabla_\Gamma \cdot \vec{u}_\Gamma $.

\begin{remark}
\label{rem:bulksurfacecons}
On account of state-function assumptions~\ref{SA1} and~\ref{SA3} 
in~\cref{subsec:postulatesassumptions}, the derivatives $G'_i$ are strictly increasing and $\eta_i^{-1}$ are strictly decreasing with respect to~$c_i$. Assumption~\ref{SA2} implies that~$\theta_i$ is strictly positive. Conservation of surfactants then implies that at each interface temperature~$\theta_{\Gamma}$, there is a unique combination of the bulk concentrations at the interface~$c_i|_\Gamma$ and the interface concentration~$c_{\Gamma}$ such that~$G'_i (c_i) - \theta_{\Gamma}/{\eta_{i}} = G_\Gamma'(c_{\Gamma})  - \theta_{\Gamma}/\eta_{\Gamma}\text{ for }i=1,2$ holds.     
\end{remark}

\begin{remark}\label{rem:LaplaceMarangoni}
The first term in the surface pressure~\eqref{eq:surfacepressure} may be recognized as the Legendre transform of the interface energy (denoted as~$\sigma_{\scriptscriptstyle G}$) and the second term as the Legendre transform of the mixture entropy (as~$\sigma_s)$. Combining Equations~\eqref{eq:surfacepressure} and~\eqref{cons_momentum_interface} results in the Young--Laplace pressure jump across the interface and the Marangoni effect along the interface. Specifically,~\eqref{eq:surfacepressure} exhibits three contributions to these normal and tangential traction jumps, viz. a surface mixture energy effect~$\sigma_{\scriptscriptstyle G}$, an entropic effect~$ \theta_\Gamma \sigma_s  $, and a thermal effect~$-  e_{\rm{th},\Gamma}$. 
\end{remark}

\subsubsection{Constitutive relations}\label{subsubsec:constitutiverelations}
Within the confines of the constitutive class~\eqref{eq:respfuncpostulate}, and with the equations of state for $\vec{u}_{\vec{\tau}}$ and $p_\Gamma$ per \eqref{eq:tangentialinterfacevelocity} and~\eqref{eq:surfacepressure}, the entropy-production inequality~\eqref{eq:balance_entropy} can only be satisfied if the individual terms are non-negative, i.e.
\begin{equation}\label{eq:bulk_flux_inequalities}
\left.\begin{aligned}
    \frac{1}{\theta_i} \vec{\tau}_{i} :  \nabla\vec{u}_i \geq 0 \,\,\,\,\Rightarrow\,\,\,\,\,  \vec{\tau}_{i} :  \nabla\vec{u}_i \geq 0 &\\
    - \frac{1}{\theta_{i}^2}\nabla\theta_{i}\cdot \vec{q}_i \geq 0 \,\,\Rightarrow\,\,\,-\nabla\theta_{i}\cdot \vec{q}_i \geq 0&\\
    - \Big( \frac{1}{\theta_i} \nabla G_{i}'(c_{i})  +\frac{1}{\eta_{i}^2} \nabla \eta_{i}\Big)\cdot \vec{J}_i \geq 0 \,\,\Rightarrow\,\,- \nabla c_{i} \cdot \vec{J}_i \geq 0&\\
\end{aligned}\right\}\text{ for } i=1,2,\Gamma\,,
\end{equation}
where the gradient operator should again be interpreted as the surface gradient operator in the case~$i=\Gamma$. The implication in~\eqref{eq:bulk_flux_inequalities} follow from the non-negativity of the temperature, and from the convexity of $G_i$ according 
to~\ref{SA1} and the concavity of $s_i$ according to~\ref{SA2}. 


Arguably the simplest cases of constitutive relations in accordance with~\eqref{eq:bulk_flux_inequalities} are the linear relations. For the 
deviatoric stress tensor, the linear relation represents a Newtonian fluid, defined as $\boldsymbol{\tau}_{i}:={^4}\mathbb{C}^{i}:\nabla\boldsymbol{u}_i$. The inequality then holds for all symmetric positive semi-definite fourth-order tensors ${^4}\mathbb{C}^{i}$. The simplest example is ${^4}\mathbb{C}^{i}=2\mu_{i}\,{^4}\mathbb{I}^{s}$ with $\mu_i> 0$ as the dynamic viscosity, which leads to
$\boldsymbol{\tau}_{i}=\mu_i((\nabla\vec{u}_i)+(\nabla\vec{u}_i)^T)$.
In combination with the compressibility-based dissipative response on the interface, this is known as the Boussinesq--Scriven model. For the heat flux and the surfactant-concentration flux, the linear relations are respectively called Fourier's law, defined as $\boldsymbol{q}_{i}:=-\mathbb{L}_{i}\cdot\nabla\theta_{i}$ and Fick's law, $\vec{J}_{i}:=-\mathbb{D}_{i}\cdot\nabla c_{i}$. Again, the inequality holds for all symmetric positive semi-definite second-order tensors $\mathbb{L}_{i}$ and $\mathbb{D}_i$, with the simplest examples being $\mathbb{L}_{i}=\lambda_i\mathbb{I}$, with $\lambda_i> 0$ the thermal conductivity, and $\mathbb{D}_{i}=D_i\mathbb{I}$, with $D_i> 0$ the diffusion constant.

\subsubsection{Interface conditions}\label{subsubsec:interfaceconditions}
The interface conditions are derived analogously to the constitutive relations. Within the confines of the constitutive class~\eqref{eq:respfuncpostulate}, Equation~\eqref{eq:balance_entropy} insists that the bulk fluxes at the interface comply with the following inequalities:
\begin{equation}\label{eq:interfaceconditionineq}
\begin{split}
\left.\begin{split}      
    (-1)^{i+1}\left(\frac{1}{\theta_{\Gamma}}-\frac{1}{\theta_{i}}\right)\vec{q}_{i}\cdot\vec{\nu}\geq 0 \,\,\,\,\Rightarrow\,\,\,\,\, (-1)^{i+1}\left(\theta_{i} - \theta_{\Gamma} \right)\vec{q}_{i}\cdot\vec{\nu}&\geq 0\\
    (-1)^{i+1}\left(-\frac{G'_{\Gamma}\left(c_{\Gamma}\right)}{\theta_{\Gamma}}+\frac{G'_{i}\left(c_{i}\right)}{\theta_{\Gamma}}+\frac{1}{\eta_{\Gamma}}-\frac{1}{\eta_{i}}\right)\vec{J}_{i}\cdot\vec{\nu}\geq0\,\,\,\,\Rightarrow\quad& \\
     (-1)^{i+1}\bigg(\bigg(G'_i(c_i)-\frac{\theta_{\Gamma}}{\eta_i}\bigg)-
     \bigg(G'_{\Gamma}(c_{\Gamma})-\frac{\theta_{\Gamma}}{\eta_{\Gamma}}\bigg)
     \bigg)\vec{J}_{i}\cdot\vec{\nu}&\geq0
\end{split}\right\}&\text{ at }\Gamma\text{ for }i=1,2, \\
    -\left(\avg{\vec{\tau}_i-p_i\vec{I}}\cdot\vec{\nu}\right)\cdot\jump{\vec{u}_{\vec{\tau},i}}\geq0\,\,\,\,\,\,& \text{ at }\Gamma\,,\\
\end{split}
\end{equation}
where the occurrence of $(-1)^{i+1}$ is due to the convention of the surface normal, and the implications in the first and second line follow from multiplication with (non-negative) $\theta_i\theta_\Gamma$ and~$\theta_\Gamma$, respectively.


The simplest interface conditions satisfying the first and third inequalities in~\eqref{eq:interfaceconditionineq} are again linear scaling relations. For the jump in the heat transfer, this is called Newton's law of cooling: $\vec{q}_i \big|_\Gamma \cdot\vec{\nu} = (-1)^{i+1} \alpha_i (\theta_i - \theta_\Gamma)\text{ }(i=1,2)$ with $\alpha_i$ as the heat transfer coefficient. For the traction jump, this represents the Navier slip condition~$ \mathbf{P}_{\Gamma}\big(\avg{\vec{\tau}_i-p_i\vec{I}}\cdot\vec{\nu}\big) = -\beta_{s}\jump{\vec{u}_{\vec{\tau},i}}$ with~$\beta_{s}\geq{}0$ as the slip coefficient.  The somewhat more convoluted form of the second inequality precludes a simple linear scaling relation. Instead, the simplest closure relation is:
\begin{align}\label{eq:surfactantadsorption}
    \vec{J}_{i}|_{\Gamma}\cdot\vec{\nu}= (-1)^{i+1}\gamma_{i}\bigg(\bigg(G'_i(c_i)-\frac{\theta_{\Gamma}}{\eta_i}\bigg)-
     \bigg(G'_{\Gamma}(c_{\Gamma})-\frac{\theta_{\Gamma}}{\eta_{\Gamma}}\bigg)
     \bigg) \,,
\end{align}
with $\gamma_i$ as the adsorption-desorption coefficient. Interface condition~\eqref{eq:surfactantadsorption} admits a convenient interpretation via the chemical potentials of the bulk and the interface; see~\cref{remark:Isotherms} below.

\newpage
\section{Equilibrium}
\label{sec:special_cases}
The characterization of equilibrium states of evolution equations with a conservation-dissipation structure, such as the presented binary-fluid-surfactant system,
forms an integral part of their analysis. In this section, we investigate the implications of the derived closure relations in this regard. Additionally, we investigate concentration distribution and surface tension in equilibrium for an example system.

\subsection{Equivalences}
\label{ssec:equivalences}
We regard the binary-fluid-surfactant system, equipped with the equations of state~\eqref{eq:tangentialinterfacevelocity} and~\eqref{eq:surfacepressure}, and the constitutive relations and interface conditions introduced in~\cref{subsubsec:constitutiverelations,subsubsec:interfaceconditions}. To characterize equilibrium solutions of the binary-fluid-surfactant system, 
we define three conditions that indentify an equilibrium state, and we establish that these conditions are equivalent. The equilibrium conditions are: \emph{(i)} entropy production vanishes uniformly in time up to exogenous terms; \emph{(ii)} the system state $q=(\vec{u},p,c,\theta,\vec{\chi})$ belongs to a certain class of equilibrium states~$\boldsymbol{Q}_{\textsc{e}}$; \emph{(iii)} the time rate of change of the state variables vanishes:
\begin{equation}\label{eq:equilibriumequivalences}
\begin{tikzcd}[column sep=small]
& \{q(t):t>0\}\subseteq\boldsymbol{Q}_{\textsc{e}} & \\
\parbox{4.2cm}{\centering $\{\mathcal{P}(q(t)):t>0\}=\{0\}$} \arrow[rr, Leftrightarrow]{} \arrow[swap, ur, Leftrightarrow]{} & & \parbox{4cm}{$\{\partial_tq(t):t>0\}=\{0\} $} \arrow[ul, Leftrightarrow]{}
\end{tikzcd}     
\end{equation}
Specifically, the first condition implies that the left member of~\eqref{eq:balance_entropy}, extended to all of~$\Dom$, i.e. with $V_i=\Dom_i$ and ~$S_{\Gamma}=\Gamma$, vanishes uniformly in time. The second condition connotes that equilibrium states assume a particular form, to be specified below. The third condition implies that the set of equilibrium states~$\boldsymbol{Q}_{\textsc{e}}$ is invariant under the dynamics of the system.

\begin{figure}[h]
    \centering
    \subfloat[Initial]{
    \includegraphics[width=0.3\textwidth]{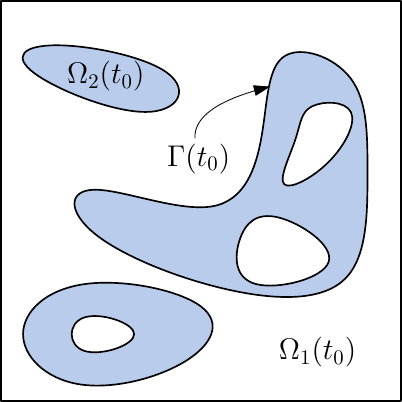}\label{fig:topoinit}}
    \hspace{0.15\textwidth}
    \subfloat[Equilibrium]{
    \includegraphics[width=0.3\textwidth]{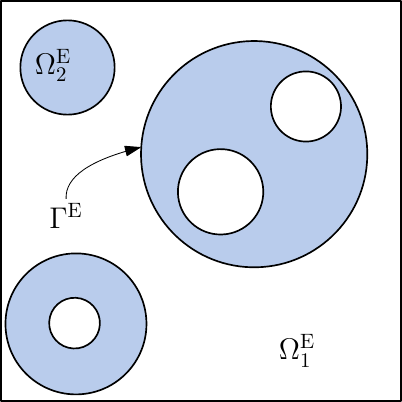}\label{fig:topoequi}}
    \caption{Schematic depiction of the initial and equilibrium topologies for nested spheres.\label{fig:topology}}    
\end{figure}

The equilibrium configuration of the binary-fluid-surfactant system depends on its initial configuration. The standing assumption that the topology of the interface does not change during the evolution of the binary-fluid-surfactant system imposes restrictions on the initial configurations and initial conditions that can be considered. For instance, it is well known that elongated fluid filaments in 3D can exhibit instabilities which cause these filaments to break up into smaller parts; see e.g.~\cite{Notz:2004tb}.
We, therefore, insist that the initial configuration of the interface corresponds to a collection of disconnected topological spheres such that this topological configuration is invariant under the evolution of the binary-fluid-surfactant system, i.e. for all $t\geq{}0$ it holds that $\Gamma(t)=\bigcup_{k=1}^{N_{\Gamma}}\Gamma_k(t)$ such that $\operatorname{dist}(\Gamma_k(t),\Gamma_l(t))>0$ for $k\neq{}l$ and each $\Gamma_k(t)$ corresponds to a topological sphere; see Figure~\ref{fig:topology} for an illustration. This condition implies that the interface does not intersect the boundary~$\partial\Omega$. Without loss of generality, we assume that $\Dom_2$ is immersed in~$\Dom_1$ in the sense that~$\partial\Omega=\partial\Omega_1\setminus\Gamma$. Each domain $\Dom_i(t)$ ($i=1,2$) is generally disconnected and comprised of~$N_i$ connected 
subsets~$\Dom_i^k(t)=\vec{\chi}(t,\tilde{\Dom}{}_i^k)$.

To specify the class of equilibrium states, we assume that the binary-fluid-surfactant system is closed by a stationary, impermeable, thermally insulated, solid boundary:
\begin{equation}
\label{eq:closedBCs}
\vec{u}_1=0,
\quad    
\vec{J}_1\cdot\vec{n}=0,
\quad
\vec{q}_1\cdot\vec{n}=0,
\qquad\text{on }\partial\Omega\,.
\end{equation}
Because~\eqref{eq:closedBCs} specifies a Dirichlet condition on the velocity on the entire external boundary~$\partial\Omega$, the pressure in the system is only defined up to a constant. To fix this constant, we insist that~$p$ vanishes on average:
\begin{equation}
\langle{}p\rangle_{\Dom}
:=
\int_{\Dom}p\d{V}
=
0\,.
\end{equation}
In conjunction with these boundary conditions and under the aforementioned conditions 
on the configuration, we characterize the class of equilibrium states as
\begin{multline}
\label{eq:QE}
\boldsymbol{Q}_{\textsc{e}}
=
\big\{
(\vec{u},p,c,\theta,\vec{\chi}):
\vec{u}_1=\vec{u}_2=\vec{u}_{\Gamma}=0,\,
p|_{\Dom^k_i}=p_{i,k}^{\textsc{e}},\,
\langle{}p\rangle_{\Dom}=0,\,
c_1=c_1^{\textsc{e}},\,
c_2=c_2^{\textsc{e}},\,
c_{\Gamma}=c_{\Gamma}^{\textsc{e}},\,
\\
\theta_1=\theta_2=\theta_{\Gamma}=\theta^{\textsc{e}},\,
\operatorname{meas}(\vec{\chi}(\tilde{\Dom}^k_i))=
\operatorname{meas}(\tilde{\Dom}^k_i),\,
\vec{\chi}(\tilde{\Gamma}_k)\text{ is spherical}\big\}
\end{multline}
for certain constants $(\cdot)^{\textsc{e}}$, subject to
\begin{equation}
\label{eq:concbal}
G'_1(c_1^{\textsc{e}})-\theta^{\textsc{e}}/\eta_1^{\textsc{e}}
=
G'_2(c_2^{\textsc{e}})-\theta^{\textsc{e}}/\eta_2^{\textsc{e}}
=
G'_{\Gamma}(c_{\Gamma}^{\textsc{e}})-\theta^{\textsc{e}}/\eta_{\Gamma}^{\textsc{e}},
\end{equation}
with $1/\eta_i^{\textsc{e}}=\partial_{c_i}s_i(c_i^{\textsc{e}})$ and
\begin{equation}
\label{eq:Young-Laplace}
p_{1,k}^{\textsc{e}}-p_{2,l}^{\textsc{e}}
=
\operatorname{sgn}\,|p_{\Gamma}^{\textsc{e}}\kappa_m|
\qquad\text{for }\tilde{\Gamma}_m=\partial\tilde{\Dom}_1^k\cap\partial\tilde{\Dom}_2^l\neq\emptyset
\end{equation}
where $p_{\Gamma}^{\textsc{e}}$ is the constant surface pressure at surface concentration~$c_{\Gamma}^{\textsc{e}}$ and temperature~$\theta^{\textsc{e}}$ according to~\eqref{eq:surfacepressure}, and $\kappa_m$ denotes the (constant) additive curvature of the sphere $\vec{\chi}(\tilde{\Gamma}_m)$. The symbol $\operatorname{sgn}$ represents~$1$
(resp.~$-1$) if $\vec{\chi}(\tilde{\Dom}{}_1^k)$ is interior (resp. exterior) to~$\vec{\chi}(\tilde{\Gamma}_m)$. One may note that~\eqref{eq:concbal} prescribes the balance of bulk and surface concentrations as indicated in~\cref{rem:bulksurfacecons}. Equation~\eqref{eq:QE} characterizes equilibrium states of the binary-fluid-surfactant system as being composed of geometrically spherical (possibly nested) droplets, immersed in an ambient fluid; the droplets are stationary, and the velocity in the binary fluid vanishes uniformly; the binary-fluid-surfactant system exhibits a homogeneous temperature;
the surfactant concentration in each of the subsystems (fluids~1 and~2, and interface) is uniform, and the bulk and interface concentrations are balanced according to~\eqref{eq:concbal}; the surface pressure is uniformly constant on the interface; the pressure is constant in each subdomain~$\Dom_i^k$ and the pressures in adjacent subdomains are related by the Young--Laplace condition.

Denoting by $\mathcal{P}(q)$ the entropy production in the left-hand side of~\eqref{eq:balance_entropy}, extended to all of~$\Dom$, we first consider the implication 
$\{\mathcal{P}(q(t)):t>{}0\}=\{0\}\Rightarrow{}\{q(t):t>0\}\subseteq\boldsymbol{Q}_{\textsc{e}}$. It is to be noted that this implication is supposed to hold subject to the evolution equations for the binary-fluid-surfactant system. Introducing
the equations of state~\eqref{eq:tangentialinterfacevelocity} and~\eqref{eq:surfacepressure}, and the constitutive relations and interface conditions introduced in~\cref{subsubsec:constitutiverelations,subsubsec:interfaceconditions}, we obtain
\begin{equation}
\label{eq:production}
\begin{aligned}
\mathcal{P}&=
\sum^{2}_{i=1} \int\limits_{\Dom_i(t)} \Big\{ \theta_i^{-1}\mu_i |\nabla\vec{u}_i|^2 + \theta_{i}^{-2}\lambda_i|\nabla\theta_{i}|^2
+D_i\big( \theta_i^{-1}\partial_{c_i}^2G_i(c_{i})-\partial_{c_i}^2s_i(e_{\mathrm{th},i},c_i)\big)|\nabla{}c_i|^2 \Big\} \d{V}   \\
&\phantom{=}
    + \int\limits_{\Gamma(t)}\!\! \Big\{ \theta_\Gamma^{-1} \mu_{\Gamma}| \nabla_\Gamma\vec{u}_\Gamma|^2  + \theta_{\Gamma}^{-2}|\nabla_{\Gamma}\theta_{\Gamma}|^2+D_{\Gamma} \big( \theta_\Gamma^{-1} \partial_{c_{\Gamma}}^2G_{\Gamma}(c_{\Gamma})  
    -\partial_{c_{\Gamma}}^2s_{\Gamma}(e_{\mathrm{th},\Gamma},c_{\Gamma})\big) |\nabla_{\Gamma}c_{\Gamma}|^2\Big\}  \d{S} \\
&\phantom{=} 
    + \int\limits_{\Gamma(t)}\!\! \theta_{\Gamma}^{-1}\beta_s\big(\jump{\vec{u}_{\vec{\tau},i}}\big)^2  \d{S} 
+\sum^{2}_{i=1}\int\limits_{\Gamma(t)}\!\!
\frac{\alpha_i}{\theta_{\Gamma}\theta_i}(\theta_i-\theta_{\Gamma})^2\d{S} \\
&\phantom{=} 
+\sum^{2}_{i=1}\int\limits_{\Gamma(t)}\!\!
\gamma_i\Big(\big(G'_i(c_i)-\theta_{\Gamma}\eta_i^{-1}\big)-
     \big(G'_{\Gamma}(c_{\Gamma})-\theta_{\Gamma}\eta_{\Gamma}^{-1}\big)
     \Big)^2\d{S} 
\end{aligned}
\end{equation}
In view of the quadratic form of the terms in~\eqref{eq:production} and the positivity of the multiplicative factors, vanishing entropy production implies:
\begin{equation}
\label{eq:grad=0}
\begin{aligned}
\nabla\vec{u}_i&=0,
\\
\nabla\theta_i&=0,
\\
\nabla{}c_i&=0
\end{aligned}
\qquad\qquad\qquad
\begin{aligned}
\nabla_{\Gamma}\vec{u}_{\Gamma}&=0,
\\
\nabla_{\Gamma}\theta_{\Gamma}&=0,
\\
\nabla_{\Gamma}c_{\Gamma}&=0,
\end{aligned}
\qquad
\begin{aligned}
\jump{\vec{u}_{\vec{\tau},i}}&=0,
\\
\theta_i&=\theta_{\Gamma},
\\
G'_i(c_i)-\theta_{\Gamma}\eta_i^{-1}
&=
G'_{\Gamma}(c_{\Gamma})-\theta_{\Gamma}\eta_{\Gamma}^{-1}.
\end{aligned}
\end{equation}
The equalities in the left (resp. center) column of~\eqref{eq:grad=0} imply 
that~$\vec{u}$, $\theta$ and $c$ are constant in the two fluid domains (resp. on the interface). The conditions $\jump{\vec{u}_{\vec{\tau},i}}=0$ and~\eqref{cons_mass_surf} and~\eqref{eq:tangentialinterfacevelocity} in turn imply 
that~$\vec{u}_1=\vec{u}_2=\vec{u}_{\Gamma}=\vec{u}^{\textsc{e}}$. The boundary condition $\vec{u}_1=0$ on~$\partial\Omega$ then implies $\vec{u}^{\textsc{e}}=0$. The equality $\theta_i=\theta_{\Gamma}$ in~\eqref{eq:grad=0} implies that the temperature is uniformly constant in~$\Dom$, i.e. $\theta_1=\theta_2=\theta_{\Gamma}=\theta^{\textsc{e}}$. The bottom-right condition in~\eqref{eq:grad=0} implies that the constant surfactant concentrations in the bulk domains and on the interface, $c_1=c_1^{\textsc{e}}$, $c_2=c_2^{\textsc{e}}$ and
$c_{\Gamma}=c_{\Gamma}^{\textsc{e}}$, satisfy~\eqref{eq:concbal}. By virtue of $\vec{u}_i=0$ (uniformly in time), the balance of linear momentum in~\eqref{cons_momentum_bulk} implies the piecewise-constant form of the bulk pressures in accordance with~\eqref{eq:QE}. Because $c_{\Gamma}$ and $\theta_{\Gamma}$ are constant, the equation of state for the surface pressure~\eqref{eq:surfacepressure} implies that~$p_{\Gamma}$ is constant, and it follows from the interface momentum balance~\eqref{cons_momentum_interface} that the subdomain pressures $p_{i,k}^{\textsc{e}}$ comply with the Young--Laplace relation~\eqref{eq:Young-Laplace}. The condition~$\langle{p}\rangle_{\Omega}=0$ that appears in~\eqref{eq:QE} is intrinsically satisfied as a constraint in the system. On account of the uniformity of the subdomain pressures and the interface pressure, the interface momentum balance~\eqref{cons_momentum_interface} moreover implies that the additive curvature~$\kappa$ is constant on each connected component of the interface, i.e. if $\tilde{\Gamma}_m=\partial\tilde{\Dom}_1^k\cap\partial\tilde{\Dom}_2^l\neq\emptyset$ then $\vec{\chi}(\tilde{\Gamma}_m)$ carries a constant additive curvature, $\kappa_m$. This implies that $\vec{\chi}(\tilde{\Gamma}_m)$ is geometrically spherical. Conservation of volume of the domain maps $\vec{\chi}_i$, according to $\partial_t\vec{\chi}_i=\tilde{\vec{u}}_i$ with $\tilde{\vec{u}}_i(t,\vec{\chi}_i^{-1}(t,\cdot)=\vec{u}(t,\cdot)$ (see Section~\ref{sec:SetNot}) and $\nabla\cdot\vec{u}_i=0$ per~\eqref{cons_mass_bulk}, implies that $\operatorname{meas}(\vec{\chi}_i(\tilde{\Dom}_i^k))=\operatorname{meas}(\tilde{\Dom}_i^k)$. Hence, we have established that $\{\mathcal{P}(q(t)):t>0\}=0\:\Rightarrow{}\{q(t):t>0\}\subseteq\boldsymbol{Q}_{\textsc{e}}$.

We next consider the implication $\{q(t):t>0\}\subseteq\boldsymbol{Q}_{\textsc{e}}\:\Rightarrow\:\{\partial_tq(t):t>0\}=\{0\}$. This implication signifies that the class of equilibrium states is invariant under the dynamics of the binary-fluid-surfactant system, comprised of the conservation laws in the bulk domains and at the interface, equipped with the constitutive relations and equations of state. First regarding the balance laws for the bulk domains~\eqref{cons_momentum_bulk}--\eqref{cons_thermalenergy_bulk}, one can infer that the right-hand sides of these equations vanish for states in~$\boldsymbol{Q}_{\textsc{e}}$. In combination with $\vec{u}=0$ in the material derivatives, it then follows that $\partial_t(\vec{u}_i,c_i,e_{\mathrm{th},i})=0$ and, because~$e_{\mathrm{th},i}$ is a strictly increasing function (of temperature) by assumption~\ref{SA3}, $\partial_t\theta_i=0$. Moreover, from~$\vec{u}_i=0$ it immediately follows 
that $\partial_t\vec{\chi}=0$. Considering next the balance laws for the 
interface~\eqref{cons_concentration_interface}--\eqref{cons_thermalenergy_interface}, one infers in a similar manner as for the bulk domains that $\partial_t(c_{\Gamma},\theta_{\Gamma})=0$. Moreover, $\partial_t\vec{u}_i=0$ in combination with~\eqref{cons_mass_surf} and the equation of state~\eqref{eq:tangentialinterfacevelocity} implies $\partial_t\vec{u}_{\Gamma}=0$. By virtue of $\partial_t(c_{\Gamma},\theta_{\Gamma})=0$, the equation of state~\eqref{eq:surfacepressure} implies $\partial_tp_{\Gamma}=0$. Finally, $\partial_tp_i=0$ follows from the previously derived relations in conjunction with~\eqref{cons_momentum_bulk} and~\eqref{cons_momentum_interface}; see~\cref{app:dtpressure}. This corroborates the implication $\{q(t):t>0\}\subseteq\boldsymbol{Q}_{\textsc{e}}\:\Rightarrow\:\{\partial_tq(t):t>0\}=\{0\}$.

Finally, we regard the implication 
$\{\partial_tq(t):t>0\}=\{0\}\:\Rightarrow\:\{\mathcal{P}(q(t)):t>0\}=\{0\}$, which completes the derivation of the equilibrium equivalences in~\eqref{eq:equilibriumequivalences}. This implication is in fact a direct consequence of the second law. The evolution equations for the binary-fluid-surfactant system imply that $d_t\Sent(q(t))=\mathcal{P}(q(t))\geq{}0$ 
with~$\Sent(q)$ the entropy of the binary-fluid-surfactant system, 
\begin{equation}
\Sent(q)=\sum^{2}_{i=1}\int\limits_{\Dom_i} s_i\big(c_i,e_{\mathrm{th},i}(\theta_i)\big) \d{V} 
+  \int\limits_{\Gamma}s_\Gamma\big(c_{\Gamma},e_{\mathrm{th},{\Gamma}}(\theta_{\Gamma})\big)\d{S}\,,
\end{equation}
and~$\mathcal{P}$ the entropy production according to~\eqref{eq:production}. For $\partial_tq(t)=0$, the chain rule yields $d_t\Sent(q(t))=0$ and, therefore, $\mathcal{P}(q(t))=0$.

\subsection{Equilibrium example}
\label{subsec:equiexamp}
To illustrate the equilibrium behavior of the presented binary-fluid-surfactant model, 
we examine the temperature dependence of the surfactant distribution and the surface tension. In this example, we assume that the volume of fluid~1 is sufficiently large in relation to the area of the interface and the volume of fluid~2, such that the surfactant concentration in fluid~1 is essentially constant irrespective of the surfactant concentration in fluid~2 and on the interface.
Per~\cref{ssec:equivalences}, the system being in equilibrium implies that the surfactant concentrations are uniform in each separate domain and the temperature is uniform throughout the entire system. We assume that the interface mass is sufficiently small to neglect its thermal entropy and thermal energy. For this example, we use the 
mixture-energy densities and configurational-entropy densities as defined in \cref{tab:entropyandenergy} and the parameters in \cref{tab:examparameters}. Fluid~1 serves as the reference level for the surfactant concentration and the energy level, i.e., the molecular energy density is zero in fluid~1. The temperature is varied in the range~$\theta^{\textsc{e}}\in[0,100]^\circ \rm C$.
\begin{table}
    \centering
    \begin{tabular}{|c|c|}
    \hline
        Quantity  & Interface \\ \hline
        \rule[0pt]{0pt}{14pt}Mixture energy & $G_{\Gamma}(c_{\Gamma})=\sigma_0+c_\Gamma\varepsilon_\Gamma$ \\
        Configurational entropy & $s_{\rm{mix,} \Gamma}(c_\Gamma)=-R \left[ c_\Gamma\log\big(\frac{c_\Gamma}{c_{\Gamma,\rm max} -c_\Gamma}\big) - c_{\Gamma,\rm max}\log\big(\frac{c_\Gamma,\rm max}{c_{\Gamma,\rm max}-c_{\Gamma}}\big)\right]$ \\ [8pt] 
        \hline
        Quantity   & Bulk \\ \hline
        \rule[0pt]{0pt}{14pt}Mixture energy & $G_{i}(c_{i})=c_i \varepsilon_i$ \\
        Configurational entropy & $s_{\rm{mix,} i}(c_i)=-R\left[c_i\log\big(\frac{c_i}{c_0}\big)-c_i\right]$ \\[8pt] 
        \hline
    \end{tabular}
    \caption{Mixture-energy density and configurational-entropy density for both the interface and bulk phases used in the model problem.}
    \label{tab:entropyandenergy}
\end{table}

\begin{table}
\caption{Parameter values used in the model problem.} \label{tab:examparameters}
\begin{tabular}{|l|c|c|}
\hline
State variables          & Symbol   & Value        \\ \hline
Gas constant & $R$ & 8.314 J/mol K \\ 
Reference surface tension  & $\sigma_0$ & 72.8 mNm \\ 
Maximum surface concentration & $ c_{\Gamma,\rm max} $ & 3.4 {\textmu}mol/m\textsuperscript{2}\\ 
Reference concentration   & $c_0$ & 2 mol/m\textsuperscript{3} \\
Molecular energy density  & $\varepsilon_1$ & 0 J/mol \\
Molecular energy density  & $\varepsilon_2$ & -5 kJ/mol  \\
Molecular energy density  & $\varepsilon_\Gamma$ & 1 kJ/mol \\
\hline
\end{tabular}
\end{table}

In equilibrium, the distribution of the surfactant over the bulk domains and the interface adheres to~\eqref{eq:concbal} with $1/\eta_i=\partial{}s_i/\partial{}c_i$. Inserting the expressions in Table~\ref{tab:entropyandenergy} in these identities leads to
\begin{equation}
\label{exampleisotherm}
\frac{c_i}{c_0}\exp\Big({\frac{\varepsilon_i-\varepsilon_\Gamma}{R\theta_\Gamma}}\Big):=c_iK_i=\frac{c_\Gamma}{c_{\Gamma,\rm max}-c_\Gamma}\,,
\end{equation}
where $K_i=(1/c_0)\exp((\varepsilon_i-\varepsilon_\Gamma)/R\theta_\Gamma)$ is the 
so-called equilibrium distribution constant, or adsorption constant, which determines the surfactant distribution between the bulk and the interface in equilibrium. Noting that $R=k_{\textsc{b}}N_{\textsc{a}}$, with $k_{\textsc{b}}$ as the Boltzmann constant and~$N_{\textsc{a}}$ as the Avogadro number, the ratio in the exponent in~$K_i$ can be conceived of as the entropy corresponding to the heat of adsorption, normalized with respect the Boltzmann constant, per mol. The equilibrium distribution constant is practically relevant in that it is a quantity that can generally be determined experimentally~\cite{Chang1995,Manikantan2020}.  An expression such as~\eqref{exampleisotherm} is commonly referred to as a surfactant isotherm; 
see~\cref{remark:Isotherms}.

\begin{figure}[h]
    \centering
    \subfloat[Surfactant concentration.]{
    \includegraphics[width=0.518\textwidth]{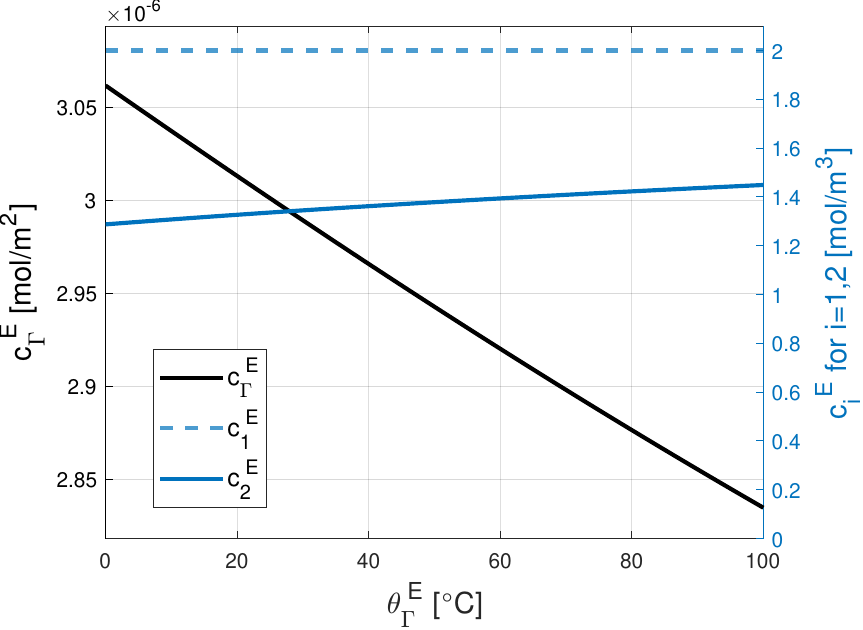}\label{fig:equiconcentration}}
    \subfloat[Surface tension.]{
    \includegraphics[width=0.482\textwidth]{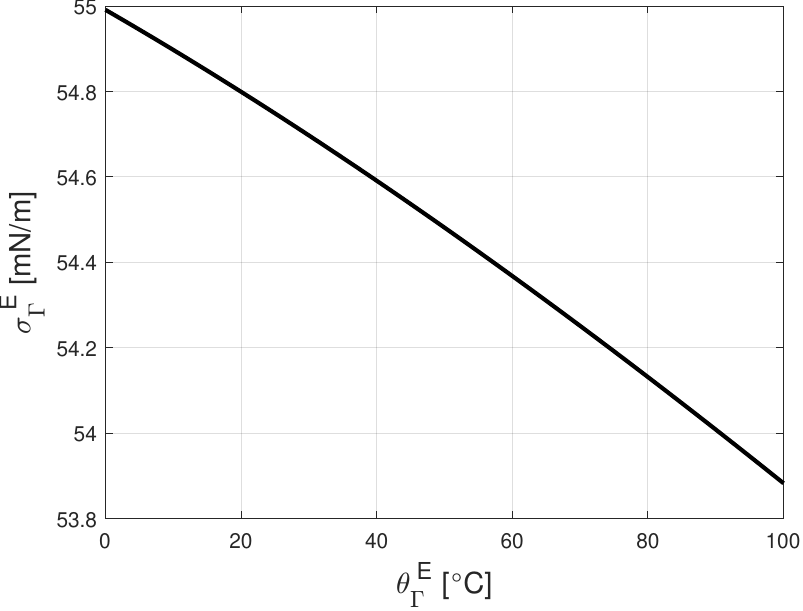}\label{fig:equisurftens}}
    \caption{Equilibrium surfactant concentrations and surface tension, as a function of temperature.\label{fig:equiX}}    
\end{figure}

Figure~\ref{fig:equiX} plots the equilibrium surfactant concentration in the bulk domains and on the interface, and the corresponding surface tension, versus temperature, for the models in Table~\ref{tab:entropyandenergy} and the parameters in Table~\ref{tab:examparameters}. The equilibrium concentrations follow from~\eqref{exampleisotherm} and, in turn, the surface tension~$\sigma_\Gamma = -p_\Gamma$ follows from~\eqref{eq:surfacepressure}. One can observe that
the equilibrium surface concentration reduces with increasing temperature because the equilibrium adsorption factor decreases. In addition, the concentration in  
fluid~1 exceeds that in fluid~2 due to the fact that the molecular-energy density of fluid~1 exceeds that of fluid~2. Panel~\ref{fig:equisurftens} conveys that the surface tension decreases with increasing temperature. To elaborate on this, we note that 
the surface tension $\sigma_{\Gamma}=-p_{\Gamma}$ exhibits a direct dependence on temperature via the term~$\theta_{\Gamma}\sigma_s$ in~\eqref{eq:surfacepressure}
and an induced dependence via the dependence of $\sigma_G$ and~$\sigma_s$ on~$c_{\Gamma}$,
and the dependence $\theta_{\Gamma}\mapsto{}c_{\Gamma}^{\textsc{e}}(\theta_{\Gamma})$ 
per~\eqref{exampleisotherm}. Hence, by the chain rule, we obtain from~\eqref{eq:surfacepressure}:
\begin{multline}
\label{eq:dsth}
\frac{\d\sigma_{\Gamma}\big(\theta_{\Gamma},c_{\Gamma}^{\textsc{e}}(\theta_{\Gamma})\big)}{\d\theta_{\Gamma}}
=
-\Big(s_{\rm{mix},\Gamma}(c_{\Gamma}^{\textsc{e}}(\theta_{\Gamma}))-c_{\Gamma}^{\textsc{e}}(\theta_{\Gamma})s'_{\rm{mix},\Gamma}(c_{\Gamma}^{\textsc{e}}(\theta_{\Gamma}))\Big)
\\
-c_{\Gamma}^{\textsc{e}}(\theta_{\Gamma})G''_{\Gamma}(c_{\Gamma}^{\textsc{e}}
(\theta_{\Gamma}))
\frac{\d{}c_{\Gamma}^{\textsc{e}}(\theta_{\Gamma})}{\d\theta_{\Gamma}}
+\theta_{\Gamma}c_{\Gamma}^{\textsc{e}}(\theta_{\Gamma})
s''_{\rm{mix},\Gamma}(c_{\Gamma}^{\textsc{e}}
(\theta_{\Gamma}))
\frac{\d{}c_{\Gamma}^{\textsc{e}}(\theta_{\Gamma})}{\d\theta_{\Gamma}}
\end{multline}
The second term vanishes on account of the linearity of~$c_{\Gamma}\mapsto{}G_{\Gamma}(c_{\Gamma})$ in~\cref{tab:entropyandenergy}. Because $s_{\rm{mix},\Gamma}$ is concave, its 2nd-order derivative is strictly negative and, due to the fact that $c_{\Gamma}^{\textsc{e}}(\theta_{\Gamma})$ is strictly decreasing (see Figure~\ref{fig:equiconcentration}), the last term in~\eqref{eq:dsth} is strictly positive. For the example functionals from \cref{tab:entropyandenergy,tab:examparameters}, the first term in~\eqref{eq:dsth} is however strictly negative and dominates the last term and, as a consequence, $\sigma_{\Gamma}$ is stricly decreasing with respect to~$\theta_{\Gamma}$. A careful observation conveys that the depicted relation between $\sigma_{\Gamma}$ and $\theta_{\Gamma}$ is slightly concave. 

\begin{remark}\label{remark:Isotherms}
    In literature, ``isotherms'' describe the surfactant distribution in a binary-fluid surfactant system in equilibrium \cite{Aveyard1973,Chang1995,Garcke2013}. Many isotherms have been formulated, such as the Henry, Langmuir, Freundlich, Volmer, and Frumkin isotherms. Any particular isotherm can be more or less suitable depending on the chemical composition of the surfactant and the concentration range of the application. These isotherms are typically derived from a postulated Helmholtz free energy ($A_i=G_i(c_i)-\theta_i s_i$), which combines energetic and entropic contributions, obtained from a partition function via statistical mechanics \cite{Aveyard1973}. The mixture energy density and configurational entropy density of \cref{tab:entropyandenergy} correspond directly to the well-known Langmuir isotherm. This isotherm is based on an interface with a monolayer of surfactant molecules with a maximum concentration of $c_{\Gamma,\rm max}$ and sufficiently low bulk concentration, i.e. below the critical micelle concentration. To show this correlation, we rewrite the surfactant adsorption 
    from~\eqref{eq:surfactantadsorption} as follows
    \begin{equation}\label{eq:adsorpchempot}
    \begin{split}
        \vec{J}_{i}|_{\Gamma}\cdot\vec{\nu} = (-1)^{i+1}\gamma_{i}\left( \mu_i(c_i)-\mu_\Gamma(c_\Gamma)\right) \\
    \end{split}
    \end{equation}
    here, $\mu_i=\pdd{}{c_i} A_i = G'_{i}(c_{i}) -\theta_\Gamma\frac{\partial s_i}{\partial c_i}  \text{ for }i=1,2,\Gamma$ is a chemical potential, defined as the partial derivative of the Helmholtz free energy density to the surfactant concentration. The isotherm then represents balance of the chemical potentials: $\mu_i(c_i)=\mu_\Gamma(c_\Gamma)$.
\end{remark}

\newpage
\section{Conclusions}\label{sec:conclusions}

We presented a systematic continuum-thermodynamics-based derivation of a binary-fluid surfactant model, comprizing two bulk fluids with an interface of co-dimension~1 and soluble surfactants in both fluids. We introduced the model and the underlying assumptions and postulates pertaining to the state functions. The considered model is based on the usual assumptions of vanishing mass density of the interface and constant mass density of the bulk phases. We derived the admissibility conditions on the constitutive relations of the model via a Coleman-Noll procedure. To this purpose, we specified the conservation relations for the binary-fluid-surfactant system, and the balance of entropy. We derived the constraints on the equations of state, constitutive relations and interface conditions imposed by the second law of thermodynamics, and we derived explicit forms for these response functions in the context of a minimal constitutive class. For the latter model, we regarded the characterization of equilibrium via vanishing entropy production, stationarity, and the particular form of equilibrium states, and proved equivalence between these different characterizations. Finally, we regarded an example with specific forms of the mixture energy and configurational entropy. In this example, we studied the surfactant concentrations and the surface tension in relation to temperature.

The considered model exhibits many classical aspects of the constitutive equations and interface conditions. The minimal constitutive class leads to a Newtonian-fluid relation for the deviatoric stress tensor, a Boussinesq--Scriven model for the interface stresses, Fourier's law for the heat flux, and Fickian diffusion of the surfactants. The combination of the interface pressure and the jump of bulk stresses over the interface, results in a stress jump with a Young--Laplace pressure and a tangential Marangoni-stress contribution. The Marangoni stress contains a thermal effect and a surfactant concentration effect. The model allows for temperature jumps over the interface, and Newton's law of cooling for the heat flux at the interface ensures thermodynamic consistency. The tangential stress at the interface and the tangential-velocity jump across the interface are connected by a Navier slip condition. 
Regarding the surfactant adsorption to the interface, thermodynamic consistency is ensured by imposing that the surfactant flux to the interface is proportional to the difference in the chemical potential between the  bulk and the interface. The surfactant adsorption to the interface hence engenders entropic and energetic effects, and the surfactant exhibits different adsorption dynamics and different equilibrium distributions depending on the state functions of the mixture energy density and the configurational entropy. In equilibrium, the heat of adsorption, i.e. the entropy generated by the reduction of the mixture energy due to transfer of surfactant from the bulk to the interface, and the configurational entropy lost by fixation of surfactants to the interface, are balanced.

For the considered binary-fluid-surfactant model, we regarded the equivalence between three different characterizations of equilibrium, viz. vanishing entropy production, stationarity of the solution, and the structure of the solution in equilibrium. We established that equilibrium is characterized by uniformity of the temperature and vanishing of the velocity of the two fluid components and of the interface. Moreover, under suitable initial conditions, the equilibrium state is composed of geometrically spherical droplets of one fluid species immersed in the other. The surfactant concentration in each fluid domain and on the interface is constant, and the surfactant concentrations are related by balance of the chemical potentials. In addition, the pressure is piecewise constant, and complies with the Young--Laplace relation. The particular equilibrium example elucidated the influence of the equilibrium temperature on the surfactant distribution and on the surface tension.

The binary-fluid-surfactant model that we have derived in this work via the Coleman-Noll procedure, is by construction thermodynamically consistent. The model can for instance be used for simulating the dynamics of binary-fluid-surfactant systems based on a numerical implementation of the model. In future work, we will focus on extending the presented model to the diffuse-interface setting. Alternate important extensions of the model could address generalizations of the constitutive class to allow for more sophisticated behavior, or consideration of wall interactions, e.g. addressing dynamic-wetting scenarios. 

\section*{Acknowledgement}
\noindent
T.B.\ van Sluijs and S.K.F.\ Stoter gratefully acknowledge the financial support through the Industrial Partnership Program {\it Fundamental Fluid Dynamics Challenges in Inkjet Printing\/} ({\it FIP\/}), a joint research program of Canon Production Printing, Eindhoven University of Technology, University of Twente, and the Netherlands Organization for Scientific Research (NWO).

\bibliographystyle{unsrtnat}
\bibliography{MyBib}

\appendix
\section{Pressure time derivative}\label{app:dtpressure}
The purpose of this section is to establish the implication $q\in\boldsymbol{Q}_{\textsc{e}}\:\Rightarrow\:\partial_tp=0$. Considering the conservation-of-momentum equation for the bulk~\eqref{cons_momentum_bulk}, and noting that $\rho_i$ is constant, we infer:
\begin{equation}
\label{eq:appA1}
    \partial_t\vec{u}_i+(\vec{u}_i\cdot\nabla)\vec{u}_i = - \frac{1}{\rho_i}\nabla p_{i} + \frac{1}{\rho_i}\nabla\cdot\vec{\tau}_{i},
\end{equation}
By substituting the constitutive relation $\vec{\tau}_i=\mu_i((\nabla\vec{u}_i)+(\nabla\vec{u}_i)^T)$ and subsequently differentiating the left and right members of equation~\eqref{eq:appA1} to~$t$, one obtains:
\begin{equation}
\partial_t^2\vec{u}_i
+
(\partial_t\vec{u}_i\cdot\nabla)\vec{u}_i
+
(\vec{u}_i\cdot\nabla)\partial_t\vec{u}_i
=
- \frac{1}{\rho_i}\nabla \partial_tp_{i}
+\frac{\mu_i}{\rho_i}
\Big(\nabla\cdot\nabla\partial_t\vec{u}_i+\nabla(\nabla\cdot\partial_t\vec{u}_i)\Big)
\end{equation}
From the prior implication $q\in\boldsymbol{Q}_{\textsc{e}}\:\Rightarrow\:\partial_t\vec{u}=0$ (see Section~\ref{ssec:equivalences}) it then follows that
\begin{equation}
 \partial_t^2\vec{u}_i = -\frac{1}{\rho_i}\nabla \partial_tp_i.
\end{equation}
and, therefore,
\begin{equation}
    \sum^{2}_{i=1}\int\limits_{\Omega_i(t)}\!\! \Ddot{\vec{u}}_i \cdot \nabla\varphi \d{V} = \sum^{2}_{i=1}-\frac{1}{\rho_i}\int\limits_{\Omega_i(t)}\!\! \nabla \partial_tp_i \cdot \nabla\varphi \d{V},
    \qquad\forall\varphi\in{}C^1(\Omega)
\end{equation}
Integration-by-parts then yields the identity:
\begin{multline}
\label{eq:varstat}
\int\limits_{\Gamma(t)}\!\! \varphi\,\partial_t^2\jump{\vec{u}_i\cdot \boldsymbol{\nu}} \d{S} 
    +\int\limits_{\partial\Omega}\!\! \varphi\,\partial_t^2(\vec{u}_1\cdot \vec{n})  \d{S} 
    -\sum^{2}_{i=1}\int\limits_{\Omega_i(t)}\!\! \varphi\, \partial_t^2(\nabla\cdot\vec{u}_i)  \d{V}    
    \\
    =\sum^{2}_{i=1} -\frac{1}{\rho_i}\int\limits_{\Omega_i(t)}\!\! \nabla \partial_tp_i \cdot \nabla\varphi \d{V}\qquad\forall\varphi\in{}C^1(\Omega).
\end{multline}
Because 
$\jump{\vec{u}_i\cdot \boldsymbol{\nu}}=0$ on account of~\eqref{cons_mass_surf},
$\vec{u}_1|_{\partial\Omega}=0$ due to~\eqref{eq:closedBCs}, and
$\nabla\cdot\vec{u}_i=0$ owing to~\eqref{cons_mass_bulk}, uniformly in time, the left-hand side of~\eqref{eq:varstat} vanishes.

We next consider the constitutive relation~\eqref{cons_momentum_interface}. By inserting the constitutive relations for the viscous-stress tensors in the bulk and on the interface, and subsequently differentiating the left and right members to~$t$, one obtains:
\begin{multline}
\label{eq:dtcons_momentum_interface}
\jump{\mu_i\big((\nabla\partial_t\vec{u}_i)+(\nabla\partial_t\vec{u}_i)^T\big)}\cdot \vec{\nu} 
- 
\jump{p_i\vec{\nu} } 
\\
= 
\nabla_\Gamma \cdot \mu_{\Gamma}\big((\nabla_{\Gamma}\partial_t\vec{u}_{\Gamma})+(\nabla_{\Gamma}\partial_t\vec{u}_{\Gamma})^T\big) 
- 
\nabla_\Gamma \partial_tp_\Gamma 
- 
\partial_t(p_\Gamma \kappa \vec{\nu})  \,. 
\end{multline}
The results in Section~\ref{ssec:equivalences} convey that for $q\in\boldsymbol{Q}_{\textsc{e}}$, it holds that $\partial_t\vec{u}_i=0$, $\partial_t\vec{u}_{\Gamma}=0$, $\partial_tp_{\Gamma}=0$ and $\partial_t(\kappa\vec{\nu})=0$. The latter is a consequence of~$\partial_t\vec{\chi}_i=0$. Equation~\eqref{eq:dtcons_momentum_interface} therefore reduces to~$\jump{\partial_t{p}_i}=0$, which implies that $\partial_tp_i$ is continuous across the interface. Under mild assumptions on the smoothness of~$p_i$ in the subdomains, $\partial_tp_i$ can be conceived of as the restriction of a function $\partial_tp\in{}H^1(\Omega)$ to~$\Dom_i$. The right-member of~\eqref{eq:varstat} then extends by continuity to~$\varphi\in{}H^1$ and, recalling that the left member of~\eqref{eq:varstat} vanishes, one obtains:
\begin{equation}
\label{eq:almostthere}
\partial_tp\in{}H^1(\Omega):
\qquad
\int\limits_{\Dom}\!\! \nabla \partial_t{p} \cdot \nabla\varphi \d{V}=0
\qquad\forall\varphi\in{}H^1(\Omega)\,.
\end{equation}
From the condition~$\langle{}p\rangle_{\Dom}=0$ associated with the boundary conditions~\eqref{eq:closedBCs}, it moreover follows that~$\langle{}\partial_tp\rangle_{\Dom}=0$ which, in conjunction with~\eqref{eq:almostthere}, implies $\partial_tp=0$.
\section{Weak formulation}\label{app:WF}

Deriving the weak formulation of a partial differential equation serves as a check for the completeness of the system description: if unbounded terms remain, some conditions are unused, or the number of unknowns is not equal to the number of test functions, then this is cause for concern. In this appendix, we derive the weak formulation for the case of the linear closure relations discussed in \cref{subsubsec:constitutiverelations,subsubsec:equationsofstate,subsubsec:interfaceconditions}. For ease of presentation, homogeneous boundary conditions are assumed along the exterior boundary.

By multiplying the conservation laws \eqref{cons_momentum_bulk},~\eqref{cons_mass_bulk},~\eqref{cons_concentration_bulk},~\eqref{cons_thermalenergy_bulk},~\eqref{cons_concentration_interface}, and~\eqref{cons_thermalenergy_interface} by testfunctions, integrating over their respective domains and integrating by parts we obtain:
{\small \begin{align}
&\text{For } i\in\{1,2,\Gamma\}, \text{ find } (\vec{u}_i ,p_i,c_i,\theta_i ) \in \mathcal{V}_i \text{ such that }\forall\, (\vec{v}_i,q_i,w_i,\xi_i ) \in \mathcal{V}_{i,0}: \nonumber\\
&\begin{cases}
\begin{split}
    &\!\!\! \sum\limits_{i=1}^2\int\limits_{\Dom_i} \Big\{  \pdd{}{t} ( \rho_i \u_i ) \cdot \vec{v}_i + \nabla \cdot( \rho_i \u_i\otimes \u_i )\cdot\vec{v}_i + \vec{\tau}_i:\nabla^s \vec{v}_i - p_i \nabla\cdot\vec{v}_i \Big\}\dDom + \\
    &\!\!\!\!\! \hspace{1cm}\int\limits_{\Gamma} \Big\{ -\big( \jump{ \vec{\tau}_i-p_i\vec{I}}\cdot\vec{\nu}\big) \cdot \big(\vec{v}_{\vec{\nu}} + \avg{\vec{v}_{i,\vec{\tau}}}\big) - \big( \avg{ \vec{\tau}_i-p_i\vec{I}}\cdot\vec{\nu}\big) \cdot \jump{\vec{v}_{i,\vec{\tau}}}   \Big\}\dbdy = 0\,, 
\end{split}\\
&\!\!\! \displaystyle \sum\limits_{i=1}^2 \int\limits_{\Dom_i} q_i \nabla\cdot \u_i \dDom  = 0\,,\\
\begin{split}
    &\!\!\! \displaystyle \sum\limits_{i=1}^2 \int\limits_{\Dom_i} \Big\{ \pdd{}{t} c_i \, w_i + \u_i \cdot \nabla c_i \, w_i  - \vec{J}_i \cdot \nabla w_i \Big\} \dDom  + \int\limits_{\Gamma} \jump{\vec{J}_i\cdot\vec{\nu} \,w_i } \dbdy = 0 \,, 
\end{split}\\ 
\begin{split}
    &\!\!\! \sum\limits_{i=1}^2 \displaystyle \int\limits_{\Dom_i} \Big\{ C_{i,V} \pdd{}{t} \theta_i \, \xi_i + C_{i,V} \u_i \cdot \nabla\, \theta_i \, \xi_i  - \vec{q}_i \cdot \nabla \xi_i  - \Sigma_i \,\xi_i\Big\} \dDom + \int\limits_{\Gamma} \jump{\vec{q}_i\cdot\vec{\nu} \, \xi_i}  \dbdy = 0 \,,
\end{split}\\ 
&\!\!\! \displaystyle \int\limits_\Gamma \Big\{ \pdd{}{t} c_\Gamma \, w_\Gamma + \nabla_\Gamma\cdot ( c_{\Gamma} \vec{u}_{\vec{\tau}} ) \, w_\Gamma -  \kappa  \, c_{\Gamma} u_{\vec{\nu}} \, w_\Gamma  - \vec{J}_\Gamma \cdot \nabla_\Gamma w_\Gamma - \jump{\vec{J}_i}\cdot\vec{\nu} \, w_\Gamma \Big\} \dbdy = 0\,,\\ 
&\!\!\! \displaystyle \int\limits_\Gamma \Big\{ \pdd{}{t} e_{\text{th},\Gamma} \, \xi_\Gamma +\nabla_\Gamma\cdot ( e_{\text{th},\Gamma} \vec{u}_{\vec{\tau}} ) \, \xi_\Gamma -  \kappa  \, e_{\text{th},\Gamma} u_{\vec{\nu}} \, \xi_\Gamma  - \vec{q}_\Gamma \cdot \nabla_\Gamma \xi_\Gamma  - \jump{\vec{q}_i}\cdot\vec{\nu}\, \xi_\Gamma -\Sigma_\Gamma\,\xi_\Gamma \Big\}  \dbdy = 0 \,.
    \end{cases} \label{WF1} \raisetag{5cm}
\end{align}}
with $C_{i,V} = \pdd{e_{i,\text{th}}}{\theta_i}$, and where the testfunction $\vec{v}$ is assumed to comply with the interface mass-balance relation $\jump{\vec{v}_{i,\vec{\nu}}}=\vec{0}$ from Equation~\eqref{cons_mass_surf}.
%


When we consider a Newtonian fluid with Fourier and Fick's law for the heat and concentration flux, with the equations of state and interface conditions as described in \cref{sec:Coleman_Noll}, all the unknown and unbounded terms can be treated by substitution of the appropriate closure relations:
{\small \begin{align}
&\text{For } i\in\{1,2,\Gamma\}, \text{ find } (\vec{u}_i ,p_i,c_i,\theta_i ) \in \mathcal{V}_i \text{ such that }\forall\, (\vec{v}_i,q_i,w_i,\xi_i ) \in \mathcal{V}_{i,0}: \nonumber\\
    &\begin{cases}
     \begin{split}
    &\!\!\! \sum\limits_{i=1}^2\int\limits_{\Dom_i} \Big\{  \pdd{}{t} ( \rho_i \u_i ) \cdot \vec{v}_i + \nabla \cdot( \rho_i \u_i\otimes \u_i )\cdot\vec{v}_i + 2 \eta_i \nabla^s \vec{u}_i :\nabla^s \vec{v}_i - p_i \nabla\cdot\vec{v}_i \Big\}\dDom +  \\[-0.1cm]
    & \hspace{1cm} \int\limits_{\Gamma} \Big\{  \big(\nabla_\Gamma ( \sigma_{\scriptscriptstyle G} + \theta_\Gamma \sigma_s -  e_{\rm{th},\Gamma}) + ( \sigma_{\scriptscriptstyle G} + \theta_\Gamma \sigma_s - e_{\rm{th},\Gamma}) \kappa \vec{\nu} \big) \cdot \big(\vec{v}_{\vec{\nu}} + \avg{\vec{v}_{i,\vec{\tau}}}\big) \\[-0.1cm]
    & \hspace{2cm} + \beta_s \, \jump{\vec{u}_{i,\vec{\tau}}}\cdot \jump{\vec{v}_{i,\vec{\tau}}} \Big\}  \dbdy = 0\,, 
\end{split}\\
&\!\!\! \displaystyle \sum\limits_{i=1}^2\int\limits_{\Dom_i} q_i \nabla\cdot \u_i \dDom  = 0\,,\\
\begin{split}
    & \!\!\! \displaystyle \sum\limits_{i=1}^2\int\limits_{\Dom_i} \Big\{ \pdd{}{t} c_i \, w_i + \u_i \cdot \nabla c_i \, w_i  + D_i\nabla c_i \cdot \nabla w_i \Big\} \dDom +\\[-0.1cm]
    & \hspace{1cm} \int\limits_{\Gamma} \jump{  \gamma_{i}\left( \frac{\eta_i \eta_\Gamma}{\theta_{\Gamma}} \big( G'_{i}(c_{i}) -  G'_{\Gamma}(c_{\Gamma}) \big) +\eta_{i}-\eta_{\Gamma}\right) w_i} \dbdy = 0\,,
\end{split}\\ 
\begin{split}
& \!\!\!\displaystyle \sum\limits_{i=1}^2\int\limits_{\Dom_i} \Big\{ C_{i,V} \pdd{}{t} \theta_i \, \xi_i + C_{i,V} \u_i \cdot \nabla \theta_i \, \xi_i + K_i\nabla\theta_i \cdot \nabla \xi_i    \\[-0.1cm]
&\hspace{1.5cm} - 2 \eta_i \nabla^s \vec{u}_i :  \nabla^s \vec{u}_i \,\xi_i - G_i''D_i \nabla c_i \cdot \nabla c_i \,\xi_i\Big\} \dDom+ \int\limits_\Gamma \jump{\alpha_i (\theta_i-\theta_\Gamma ) \xi_i } \dbdy = 0 \,,
\end{split} \\
\begin{split}
& \displaystyle \int\limits_\Gamma \Big\{ \pdd{}{t} c_\Gamma \, w_\Gamma + \nabla_\Gamma\cdot ( c_{\Gamma} \avg{ \vec{u}_{i,\vec{\tau}} } ) \, w_\Gamma -  \kappa  \, c_{\Gamma} u_{\vec{\nu}} \, w_\Gamma  + D_\Gamma \nabla_\Gamma c_\Gamma\cdot \nabla_\Gamma w_\Gamma \\[-0.1cm]
&\hspace{1.5cm} - \jump{  \gamma_{i}\left(\frac{\eta_i \eta_\Gamma}{\theta_{\Gamma}} \big( G'_{i}(c_{i}) -  G'_{\Gamma}(c_{\Gamma}) \big) +\eta_{i}-\eta_{\Gamma}\right) } \, w_\Gamma \Big\} \dbdy = 0\,,
\end{split}\\[0.9cm]
\begin{split}
& \displaystyle \int\limits_\Gamma \Big\{ C_\Gamma \pdd{}{t} \theta_\Gamma \, \xi_\Gamma  + \theta_\Gamma \sigma_s \,  \nabla_\Gamma\cdot \vec{u}_{\Gamma}  \,\xi_\Gamma + e_{\text{th},\Gamma} u_{\vec{\nu}} \kappa\,\xi_\Gamma  +K_\gamma \nabla\theta_\Gamma  \cdot \nabla_\Gamma \xi_\Gamma -  \jump{\alpha_i (\theta_i-\theta_\Gamma ) }\, \xi_\Gamma   \\[-0.1cm]
&\hspace{1.5cm}   - \jump{ \gamma_i \big( G'_i(c_i) - G_\Gamma'(c_{\Gamma}) \big)\left(\frac{\eta_i \eta_\Gamma}{\theta_{\Gamma}} \big( G'_{i}(c_{i}) -  G'_{\Gamma}(c_{\Gamma}) \big) +\eta_{i}-\eta_{\Gamma}\right)  } \,\xi_\Gamma  \\[0.3cm]
&\hspace{2.5cm}  - G_\Gamma'' D_\Gamma  \nabla_\Gamma c_{\Gamma} \cdot \nabla c_{\Gamma}\,\xi_\Gamma -\beta_s \, \jump{\vec{u}_{i,\vec{\tau}}}\cdot \jump{\vec{u}_{i,\vec{\tau}}}\,\xi_\Gamma \Big\}  \dbdy = 0 \,.
\end{split}
    \end{cases}
\end{align}}
All aforementioned closure relationships have found their way into the weak formulation, and all the unbounded interface terms in Equation~\eqref{WF1} have been treated. This points to the completeness of the description of the physical system.
\newpage
\end{document}